\documentclass[structabstract]{aa} 
\usepackage{lscape}
\usepackage[version=3]{mhchem}
\usepackage{threeparttable} 
\usepackage{enumitem} 
\usepackage{graphicx}
\usepackage{natbib}
\usepackage{color}
 
\usepackage{graphicx}
 \usepackage[caption=false]{subfig} 
\usepackage{color,soul} 

\bibpunct{(}{)}{;}{a}{}{,}
\begin{document}

  \title{History of the solar-type protostar IRAS16293--2422 as
    told by the cyanopolyynes}


   \author{A. A. Jaber
          \inst{1,}\inst{2}
          \and  C. Ceccarelli \inst{1}
          \and  C. Kahane\inst{1}
          \and S. Viti \inst{3}
          \and N. Balucani\inst{4,}\inst{1}
          \and E. Caux\inst{5,}\inst{6}
          \and A. Faure\inst{1}
          \and B. Lefloch\inst{1}
          \and F. Lique\inst{7}
          \and E. Mendoza\inst{8}
          \and D. Quenard\inst{5,}\inst{6} 
          \and L. Wiesenfeld\inst{1}}

   \institute{Univ. Grenoble Alpes, CNRS, IPAG, F-38000 Grenoble, France
              \email{ali.al-edhari@univ-grenoble-alpes.fr}
         \and
         University of AL-Muthanna, College of Science, Physics Department, 
             AL-Muthanna, Iraq
         \and
         University College London, Gower Street, London, UK
         \and
         Dipartimento di Chimica, Biologia e Biotecnologie, Perugia, Italy
         \and
         Universit\'{e} de Toulouse, UPS-OMP, IRAP, Toulouse,France
         \and
         CNRS, IRAP, 9 Av. Colonel Roche, BP 44346, 31028 Toulouse Cedex 4, France
         \and
          LOMC – UMR 6294, CNRS-Université du Havre, 25 rue Philippe Lebon, BP 1123, 
          76063   Le Havre, France 
          \and          
           Instituto de Astronomia, Geofísica e Ciencias Atmosféricas, Universidade de Sao
Paulo, Sao Paulo 05508-090, SP, Brazil}

   \date{Received ; accepted }

 
  \abstract
  {Cyanopolyynes are chains of carbon atoms with an atom of hydrogen
    and a CN group on either side. They are detected almost
    everywhere in the interstellar medium (ISM), as well as in comets. }
  { We present an extensive study of the cyanopolyynes distribution in the solar-type protostar IRAS16293-2422. The goals are (i) to obtain a
    census of the cyanopolyynes in this source and of
    their isotopologues; (ii) to derive how their abundance varies
    across the protostar envelope; and (iii) to obtain constraints on the
    history of IRAS16293-2422 by comparing the observations with the
    predictions of a chemical model. }
  {We analysed the data from the IRAM-30m  spectral survey towards IRAS16293-2422.
   The derived (SLED) of each detected cyanopolyyne was compared with the predictions from
    the radiative transfer code (GRAPES) to derive the cyanopolyyne abundances across the envelope of
    IRAS16293-2422. Finally, the derived abundances were
    compared with the predictions of the chemical model UCL\_CHEM.}
  {We detect several lines from cyanoacetylene (HC$_3$N) and
    cyanodiacetylene (HC$_5$N), and report the first detection of
    deuterated cyanoacetylene, DC$_3$N, in a solar-type protostar. We
    found that the HC$_3$N abundance is roughly
    constant ($\sim 1.3\times10^{-11}$) in the outer cold envelope of
    IRAS16293-2422, and it increases by about a factor 100 in the inner
    region where the dust temperature exceeds 80 K, namely when the
    volcano ice desorption is predicted to occur. The HC$_5$N has an
    abundance similar to HC$_3$N in the outer envelope and about a
    factor of ten lower in the inner region. The HC$_3$N abundance derived
    in the inner region, and where the increase occurs, also provide strong
    constraints on the time taken for the dust to warm up to 80 K,
    which has to be shorter than $\sim 10^3-10^4$ yr. Finally, the
    cyanoacetylene deuteration is about 50\% in the outer envelope and
    $\leq 5$\% in the warm inner region. The relatively low
    deuteration in the warm region suggests that 
    we are witnessing a fossil of the HC$_3$N abundantly
     formed in the tenuous phase of the pre-collapse and then frozen
    into the grain mantles at a later phase.}
  {The accurate analysis of the cyanopolyynes in IRAS16293-2422
    unveils an important part of its past story. It tells us
    that IRAS16293-2422 underwent a relatively fast ($\leq 10^5$ yr)
    collapse and a very fast ($\leq 10^3-10^4$ yr) warming up of the
    cold material to 80 K.}


   \maketitle
%

\section{Introduction}\label{intro}
Cyanopolyynes, H-C$_{2n}$-CN, are linear chains of $2n$ carbons bonded
at the two extremities with a hydrogen atom and a CN group,
respectively. They seem to be ubiquitous in the interstellar
medium (ISM), as they have been
detected in various environments, from molecular clouds to late-type
carbon-rich asymptotic giant branch (AGB) stars. The detection of the largest cyanopolyyne,
HC$_{11}$N, has been reported in the C-rich AGB star IRC+10216
\citep{cer96} and the molecular cloud TMC-1 \citep{be97}, which shows
an anomalously large abundance of cyanopolyynes with respect to other
molecular clouds. Curiously enough, in star-forming regions, only
relatively short chains have been reported in the literature so far,
up to HC$_7$N, in dense cold cores and warm carbon-chain
chemistry (WCCC) sources \citep[e.g.][]{sa08,co12,fr13}.
For many years, cyanopolyynes have been suspected to be possible steps
in the synthesis of simple amino acids \citep[e.g.][]{bra98}.  More
recently, the rich N-chemistry leading to large cyanopolyynes chains
observed in Titan has renewed the interest in this family of
molecules, as Titan is claimed to be a possible analogue of the early
Earth \citep{lun09}.
An important property of cyanopolyynes that is particularly relevant for
astrobiological purposes is that they are more stable and robust
against the harsh interstellar environment compared to their monomer,
and hence they may better resist the exposure to UV and cosmic
rays~\citep{Clarke:1995}. Very recent observations towards the comet
67P/Churyumov-Gerasimenko by Rosetta seem to support the idea that
cyanide polymers, and, by analogy, cyanopolyynes, are
abundant on the comet's surface \citep{goe15}.

Regardless of the role of cyanopolyynes in prebiotic chemistry, when
trapped in interstellar ices, they can certainly carry large quantities
of carbon atoms. It is, therefore, of interest to understand in detail
their formation, carbon chain accretion, and evolution in solar-like
star-forming regions.
The goal of this article is to provide the first census of
cyanopolyynes in a solar-like protostar. To this end, we used the
3-1mm unbiased spectral survey TIMASSS (Sect. \ref{sec:data-set}) towards
the well-studied protostar IRAS16293-2422 (Sect. \ref{source}), to derive
the abundance across the envelope and hot corino of HC$_3$N, HC$_5$N
and their respective isotopologues (Sect. \ref{sec:modeling}). The
measured abundance profiles provide us with constraints on the
formation routes of these species (Sect.
\ref{sec:chemical-modeling}). Section 6 discusses the implications of
the analysis, and Sect. 7 summarises our conclusions.

\begin{figure*}[bt]
  \centering
  \includegraphics[width=14cm,angle=90]{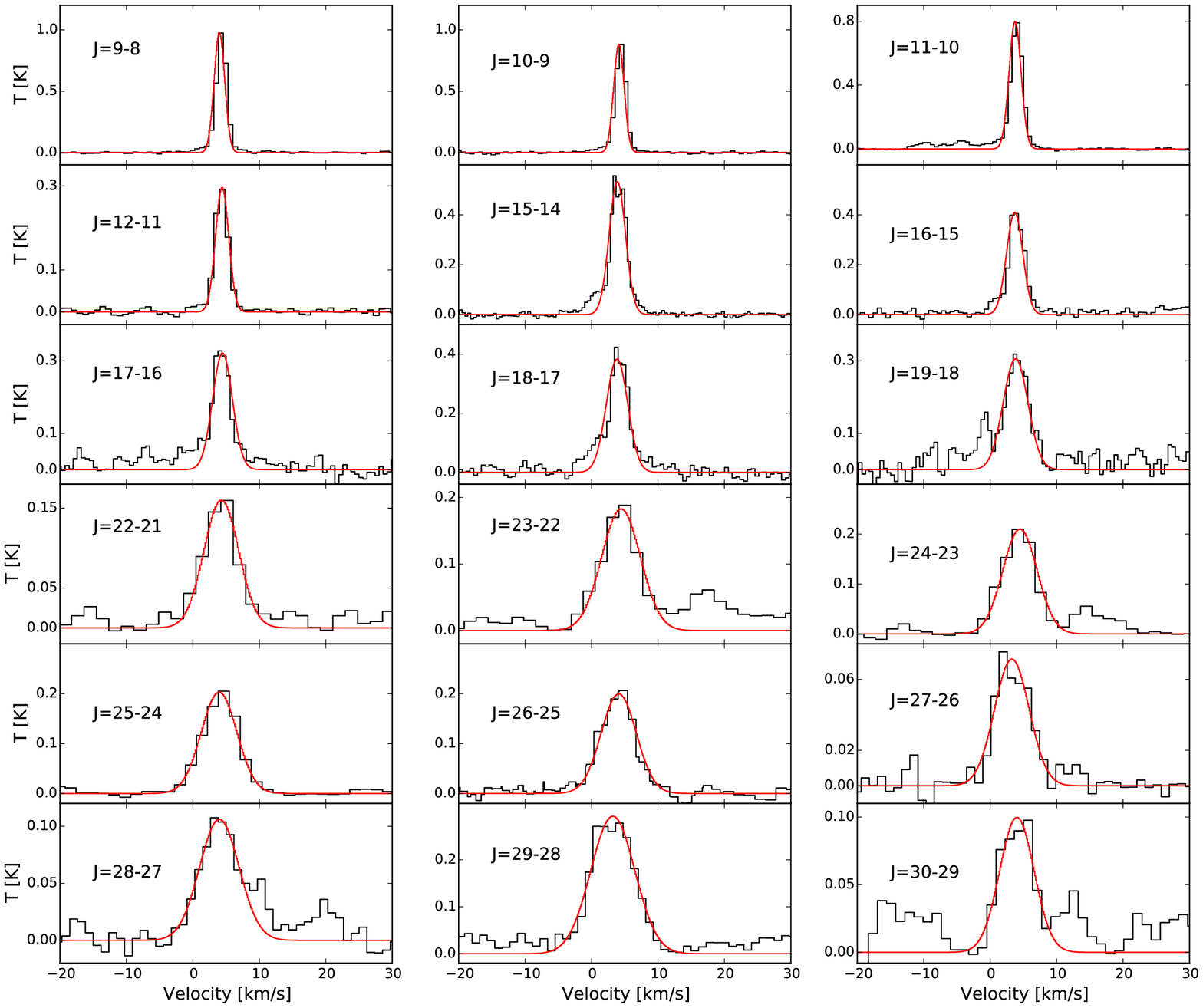}
  \caption{Observed spectra of the detected lines of HC$_3$N. The red curves
    show the Gaussian fits. The temperature is a main-beam antenna temperature.}
  \label{fig:hc3n_18}
\end{figure*}

\begin{figure*}[tb]
  \centering
 \includegraphics[width=13cm,angle=-90]{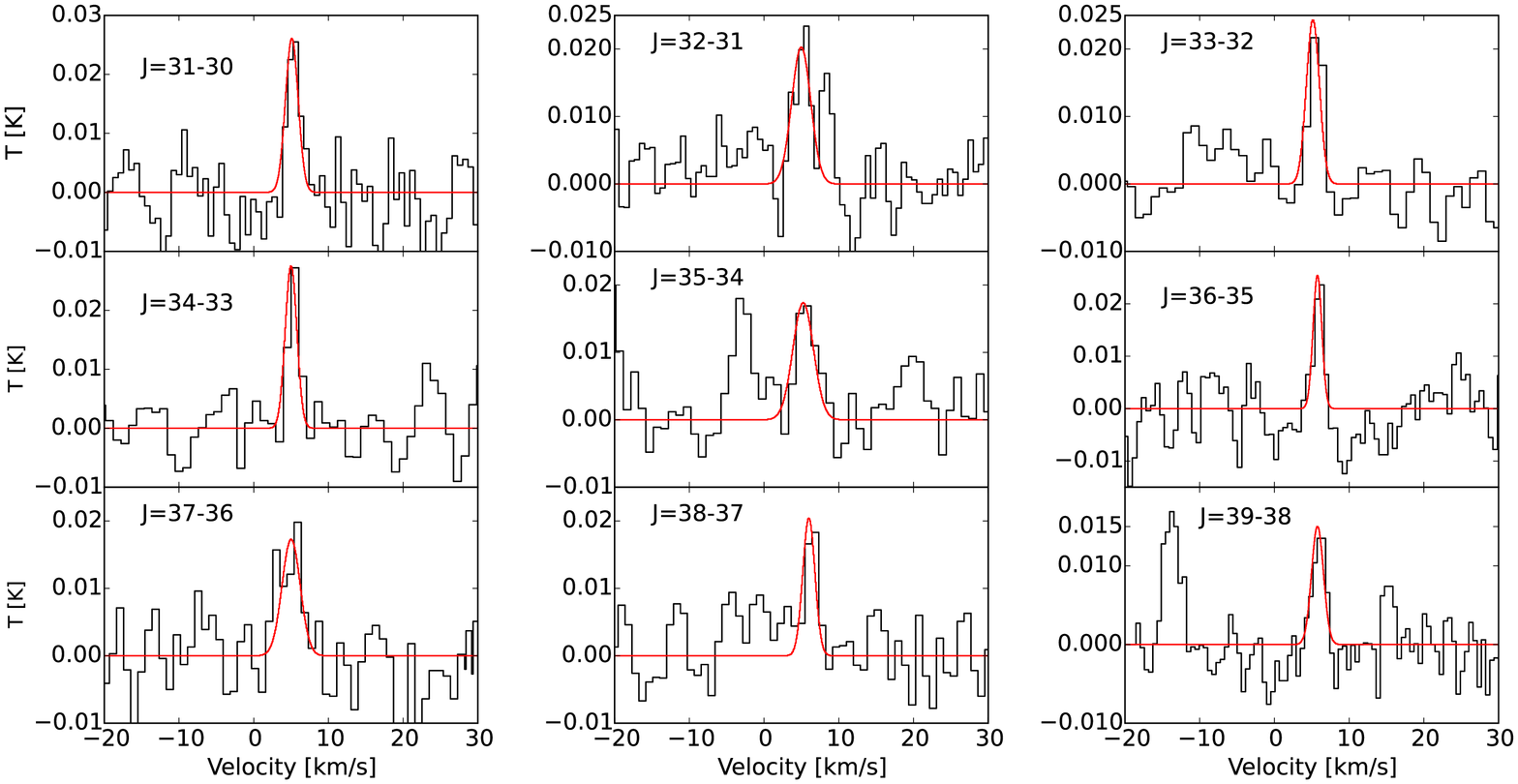}
  \caption{Observed spectra of the detected lines of HC$_5$N. The red curves
    show the Gaussian fits. The temperature is a main-beam antenna temperature.}
  \label{fig:hc5n_9}
\end{figure*}

\begin{figure*}[tb]
  \centering
 \includegraphics[width=12cm,angle=-90]{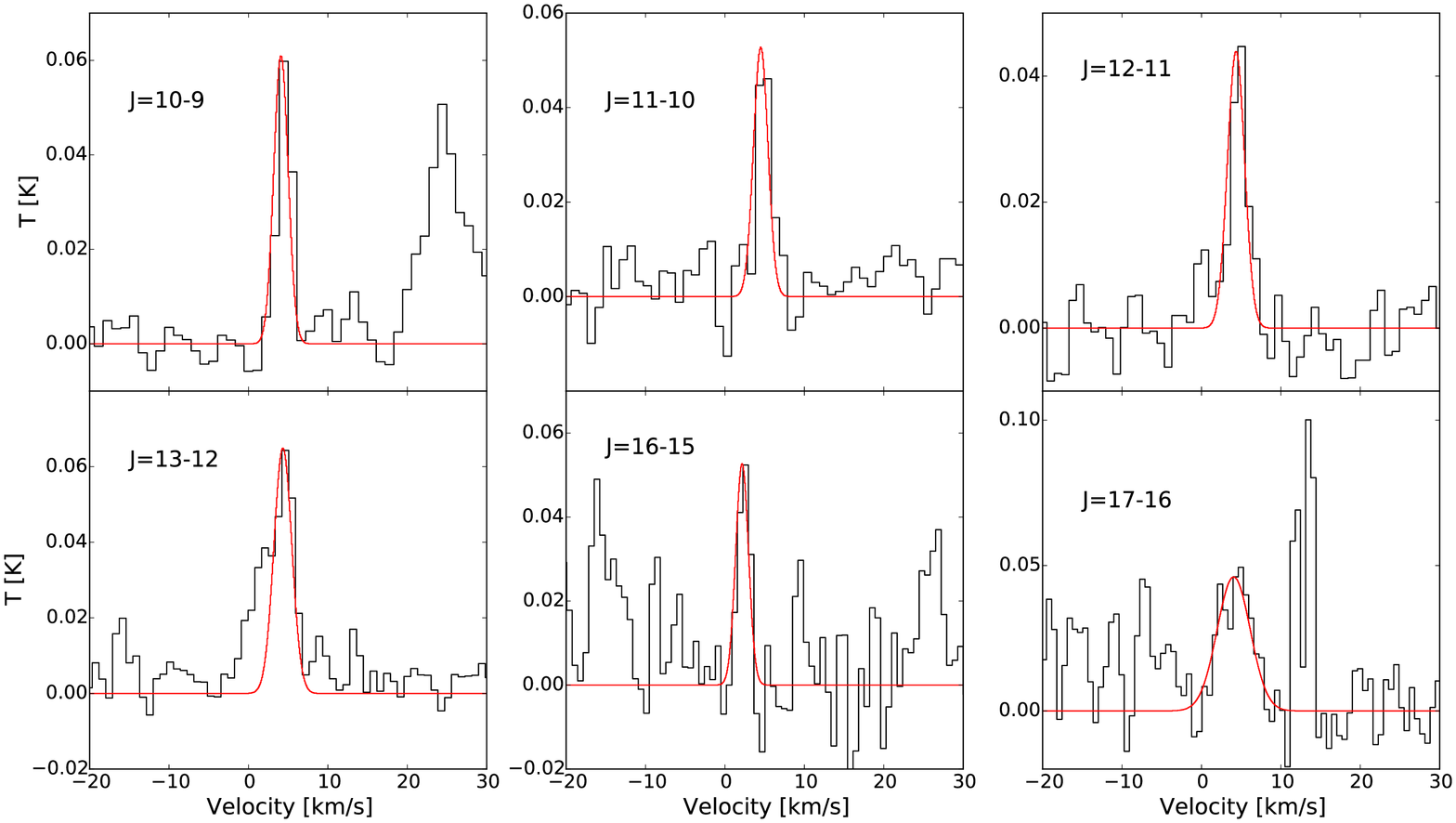}
  \caption{Observed spectra of the detected lines of DC$_3$N. The red curves
    show the Gaussian fits. The temperature is a main-beam antenna temperature.}
  \label{fig:dc3n_3}
\end{figure*}


\section{Source description } \label{source} 
IRAS16293-2422 (hereafter IRAS16293) is a solar-type Class 0
protostar in the $\rho$ Ophiuchus star-forming region, at a distance
of 120 pc \citep{lo08}.  It has a bolometric luminosity of 22
L$_{\sun}$ \citep{cr10}.  Given its proximity and brightness, it has
been the target of numerous studies that have reconstructed its
physical and chemical structure. Briefly, IRAS16293 has a large
envelope that extends up to $\sim$6000 AU and that surrounds two
sources, named I16293-A and I16293-B in the literature, separated by
$\sim5\arcsec$ \citep[$\sim$600 AU;][]{wo89,mu92}. I16293-A sizes are
$\sim1\arcsec$, whereas I16293-B is unresolved at a scale of $\sim 0.4\arcsec$
\citep{za13}.  I16293-A itself is composed of at least two sources,
each one emitting a molecular outflow \citep{mi90,lo13}.  I16293-B
possesses a very compact outflow \citep{lo13} and is surrounded by
infalling gas \citep{pi12,za13}.
From a chemical point of view, IRAS16293 can be considered as composed
of an outer envelope, characterised by low molecular abundances, and a
hot corino, where the abundance of many molecules increases by orders
of magnitude \citep[e.g.][]{cec00,sc02,caz03,ja14}.  The transition
between the two regions is presumed to occur at $\sim$100 K, which is the
sublimation temperature of the icy grain mantles.
%

\section{Data set } \label{sec:data-set}

\subsection{Observations} \label{obs}

We used data from The IRAS16293 Millimeter And Submillimeter Spectral
Survey \citep[TIMASSS;][]{ca11}.  Briefly, the survey covers the
80-280 GHz frequency interval and it has been obtained at the
IRAM-30m during the period January 2004 to August 2006 ($\sim$ 200
hr). Details on the data reduction and calibration can be found in
\citet{ca11}. We recall here the main features that are relevant for this
work. The telescope beam depends on the frequency and varies between
9$"$ and 30$"$.  The spectral resolution varies between 0.3 and 1.25
MHz, corresponding to velocity resolutions between 0.51 and 2.25 km/s.
The achieved rms ranges from 4 to 17 mK. We note that it is given in a
1.5 km/s bin for observations taken with a velocity resolution
$\leq$1.5km/s, and in the resolution bin for higher velocity
resolutions. The observations are centered on I16293-B at
$\alpha$(2000.0) = 16$^h$ 32$^m$ 22$^s$.6 and $\delta$(2000.0)=
-24$\degr$ 28$\arcmin$ 33$\farcs$. We note that the I16293-A and I16293-B
components are both inside the beam of the observations at all
frequencies.


\subsection{Species identification} \label{subsec:sp_iden}

We searched for the lines of cyanopolyynes and their isotopologues
using the spectroscopic databases Jet Propulsion Laboratory \citep[JPL;][]{pi98} and the Cologne Database for Molecular Spectroscopy \citep[CDMS;][]{mu05}. We then used the package CASSIS (Centre
d'Analyse Scientifique de Spectres Instrumentaux et Synthétiques)
({\it http://cassis.irap.omp.eu}) to fit a Gaussian to the lines.  For
the line identification we adopted the following criteria:
\begin{enumerate}
\item The line is detected with more than 3$\sigma$ in the integrated line intensity.
\item The line is not blended with other molecular lines.
\item The line intensity is compatible with the spectral line energy
distribution (SLED): since cyanopolyyynes are linear molecules,
  their SLED is a smooth curve, so that lines are discarded when their intensities  are out of the SLED defined by the majority of the lines, as shown in Fig.~\ref{fig:hc3n_compatible}.
\item The full width at half-maximum (FWHM) of the line is similar to that of the other cyanopolyyne
  lines with a similar upper level energy. In practice, {\it \textup{a
    posteriori}} the FWHM is about 2 km/s for lines with J$\leq$12 and
  increases to 6--7 km/s for J=22--30 lines.
\item The line rest velocity V$_{LRS}$ does not differ by more than 0.5 km/s
  from the V$_{LRS}$ of  the other cyanopolyyne lines.
\end{enumerate} 

The first two criteria lead to the detection of 18 lines of
HC$_3$N (from J=9-8 to J=30-29), 9 lines of HC$_5$N (from J=31-30 to
J=39-38), and the detection of 6 lines of DC$_3$N (from
J=10-9 to J=17-16). The full list of detected lines with their
spectroscopic parameters is reported in Table~\ref{tab:all}. All the
spectra are shown in Figs.~\ref{fig:hc3n_18}, \ref{fig:hc5n_9}, and
\ref{fig:dc3n_3} for HC$_3$N, HC$_5$N, and DC$_3$N, respectively. We
did not detect any line from the $^{13}$C isotopologues of HC$_3$N.

We applied the remaining three criteria and then discarded
four HC$_3$N lines (at 109.173, 245.606, 254.699, and 263.792
GHz) that had calibration problems (Caux et al. 2011) and/or
were blended with unidentified species (see
Fig.~\ref{fig:hc3n_compatible}).  Similarly, one HC$_5$N line (at
95.850 GHz) was discarded because of its abnormal values of
the intensity, FWHM, and velocity compared to the rest of the lines
(see Table ~\ref{tab:all}). Finally, one DC$_3$N (at 135.083 GHz) is
considered as tentatively detected.  

In summary, we clearly detected HC$_3$N, HC$_5$N, and DC$_3$N, for
which 14, 8, and 6 lines were identified following the five
criteria above. We did not detect any $^{13}$C isotopologue of
HC$_3$N, nor cyanopolyynes larger than HC$_5$N.

\begin{table*}[htb]
  \centering
  \caption{Parameters of the detected cyanopolyynes lines.}
  \begin{tabular}{lcccccc}
    \hline
    Transition & Frequency$^c$  & E$_{up}$ & V$_{LSR}$ & FWHM & Int. & beam \\
                   &   [MHz]      & [K]         & [km/s]     & [km/s] &[K.km/s]& ["]\\
    \hline \hline
\multicolumn{6}{c}{HC$_3$N}\\ \hline
    9-8$^a$      & 81881.4  & 19.6    & 3.9(0.6)   &2.2(0.6)  & 2.5(0.4) & 29.5 \\
    10-9$^a$     & 90979.0  & 24.0    & 3.9(0.5)   &2.1(0.5)  & 2.3(0.4) & 26.5 \\
    11-10$^a$    & 100076.3 & 28.8    & 3.9(0.4)   &2.2(0.4)  & 2.2(0.3) & 24.1 \\
    12-11$^b$    & 109173.6 & 34.1    & 3.9(0.4)   &2.3(0.4)  & 0.9(0.1) & 22.1 \\
    15-14        & 136464.4 & 52.4    & 3.9(0.3)   &2.9(0.3)  & 2.1(0.3) & 17.7  \\
    16-15        & 145560.9 & 59.4    & 3.6(0.2)   &2.8(0.3)  & 1.7(0.3) & 16.6  \\
    17-16        & 154657.2 & 66.8    & 3.9(0.2)   &3.1(0.3)  & 1.6(0.2) & 15.6 \\
    18-17        & 163753.3 & 74.7    & 3.9(0.2)   &3.5(0.3)  & 2.1(0.3) & 14.7   \\
    19-18        & 172849.3 & 82.9    & 3.8(0.2)   &4.4(0.2)  & 2.1(0.3) & 14.0 \\
    22-21        & 200135.3 & 110.5   & 3.2(0.5)   &6.1(0.7)  & 1.6(0.2) & 12.1 \\
    23-22        & 209230.2 & 120.5   & 3.4(0.5)   &6.5(0.6)  & 2.1(0.3) & 11.5 \\
    24-23        & 218324.7 & 131.0   & 3.4(0.5)   &6.1(0.5)  & 2.2(0.3) & 11.1 \\
    25-24        & 227418.9 & 141.9   & 3.4(0.4)   &6.3(0.5)  & 2.3(0.3) & 10.6 \\
    26-25        & 236512.7 & 153.2   & 3.4(0.4)   &6.2(0.5)  & 2.3(0.3) & 10.2 \\
    27-26$^b$    & 245606.3 & 165.0   & 2.7(0.5)   &6.1(0.7)  & 0.8(0.1) & 9.8  \\
    28-27$^b$    & 254699.5 & 177.2   & 3.4(0.4)   &7.0(0.4)  &1.5(0.2)  & 9.5  \\
    29-28$^b$    & 263792.3 & 189.9   & 2.5(0.3)   &7.8(0.5)  &4.7(0.7)  & 9.1  \\      
    30-29        & 272884.7 & 203.0   & 3.4(0.6)   &6.6(1.0)  & 1.4(0.2) & 8.8 \\
    \hline 
\multicolumn{6}{c}{HC$_5$N$^c$}\\ \hline
    31-30        & 82539.0  & 63.4    &3.7(0.7)    &  1.7(0.7) & 0.08(0.01) & 29.2   \\
    32-31        & 85201.3  & 67.5    &3.9(0.6)    &  2.6(0.7) & 0.07(0.01)  & 28.3   \\
    33-32        & 87863.6  & 71.7    &4.0(0.6)    &  2.0(0.6) & 0.06(0.01)  & 27.5   \\
    34-33        & 90525.9  & 76.0    &3.9(0.6)    &  1.8(0.6) & 0.06(0.01)  & 26.7   \\
    35-34        & 93188.1  & 80.5    &3.9(0.5)    &  2.7(0.5) & 0.06(0.01)  & 25.9   \\
    36-35$^b$    & 95850.3  & 85.1    &4.5(0.6)    &  1.4(0.7) & 0.04(0.01)  & 25.2   \\
    37-36        & 98512.5  & 89.8    &3.8(0.6)    &  2.7(1.2) & 0.06(0.01)  & 24.5   \\ 
    38-37        & 101174.7 & 94.7    &4.6(0.5)    &  1.8(0.6) & 0.05(0.01)  & 23.9   \\
    39-38        & 103836.8 & 99.6    &4.3(0.5)    &  1.9(0.5) & 0.03(0.01)  & 23.2   \\
    \hline 
\multicolumn{6}{c}{DC$_3$N}\\ \hline
    10-9         & 84429.8   & 22.3   &  3.9(0.6)   & 2.1(0.6)  & 0.17(0.06)     & 28.6  \\
    11-10        & 92872.4   & 26.7   &  4.4(0.5)   & 2.1(0.7)  & 0.14(0.05)     & 26.0  \\
    12-11        & 101314.8  & 31.6   &  4.3(0.4)   & 2.6(0.5)  & 0.15(0.05)     & 23.8   \\
    13-12        & 109757.1  & 36.9   & 4.1(0.4)   & 2.7(0.5)  & 0.26(0.10)     & 22.1   \\
    16-15$^d$  & 135083.1  & 55.1   &   2.1(0.3)   & 1.7(0.4)  & 0.13(0.05)     & 17.7   \\
    17-16        & 143524.8  & 62.0   & 4.3(0.2)   & 3.8(0.4)  & 0.27(0.10)     & 16.6   \\
    \hline 
  \end{tabular}
  \begin{tablenotes} 
  \item The first three columns report the transition, frequency, and
    upper energy level (E$_{up}$). The next three columns report the
    result of the Gaussian fitting: V$_{LSR}$ velocity, FWHM, and velocity-integrated line intensity
    (Int.). The last column reports the telescope beam at the
    frequency of the line.
\item $^a$: Discarded line because it is probably contaminated by the
  molecular cloud (see text, Sect. \ref{sec:results}).
\item $^b$: Discarded line because it does not satisfy criteria 3 to
  5 of Sect. \ref{subsec:sp_iden}.
\item $^c$ The HC$_3$N and DC$_3$N line frequencies are the same in
  the CDMS and JPL database. In contrast, the JPL HC$_5$N line
  frequencies are systematically overestimated with respect to those
  of CDMS, giving V$_{LSR}$ shifted by $\sim 1$ km/s with respect to
  the V$_{LSR}$ derived for the HC$_3$N and DC$_3$N lines. We
therefore  used the HC$_5$N line frequencies from the CDMS database.
\item $^d$We consider the detection of this line to be tentative.
  \label{tab:all}
  \end{tablenotes}
\end{table*}

\section{Line modelling}\label{sec:modeling}

\subsection{Model description} 
As in our previous work \citep{ja14}, we used the package GRAPES
(GRenoble Analysis of Protostellar Envelope Spectra), which is
based on the
code described in \cite{ce96,ce03}, to interpret the SLED of the detected cyanopolyynes. Briefly,
GRAPES computes the SLED from a spherical infalling envelope for a
given density and temperature structure. The code uses the beta escape
probability formalism to locally solve the level population
statistical equilibrium equations and consistently computes each line
optical depth by integrating it over the solid angle at each point of
the envelope. The predicted line flux is then integrated over the
whole envelope after convolution with the telescope beam.  The
abundance $X$ (with respect to H$_2$) of the considered species is
assumed to vary as a function of the radius with a power law $\alpha$
in the cold part of the envelope, and as a jump to a new abundance
in the warm part, corresponding to the sublimation of ices
\citep[e.g.][]{Caselli:2012}.  The transition between the two regions
is set by the dust temperature, to simulate the sublimation of the ice
mantles, and occurs at $T_{jump}$.

The abundance is therefore given by%
\begin{eqnarray}
  X(r)& = & X_{out} \left( \frac{r}{R_{max}} \right)  ^\alpha~~~~T_{dust} \le T_{jump}
\nonumber  \\ ~~
  X(r)& = & X_{in}  ~~~~~~~~~~~~~~~~~~~~~T_{dust}> T_{jump}
\label{eq:1}
,\end{eqnarray}
where $R_{max}$ represents the largest radius of the envelope.  We here used the physical structure of the IRAS16293 envelope
derived by \cite{cr10}, where the maximum radius is $1\times10^{17}$
cm.

We carried out non-local thermal equilibrium (NLTE) calculations for all detected species. We used the collisional coefficients for HC$_3$N computed
by~\cite{Faure:2016}.  For DC$_3$N collisional coefficients, we assumed the same as those of HC$_3$N. Finally, the collisional
coefficients for HC$_5$N have been computed by Lique et. al. (in
preparation).  Briefly, HC$_5$N rate coefficients were extrapolated
from HCN \citep{Ben-Abdallah:2012} and HC$_3$N \citep{we07}
considering that the cyanopolyyne rate coefficients are proportional
to the size of the molecules as first suggested by \cite{Snell:1981}.
However, to improve the accuracy of the estimation, we
considered scaling factors that also depend on the transition and the
temperature.
Hence, the ratio of HC$_3$N to HCN rate coefficients was also used to evaluate the HC$_5$N coefficients as follows:\\
\begin{eqnarray}
 k_{HC_5N}(T) = k_{HC_3N}(T) * [\frac{1}{2} + \frac{1}{2} \frac{k_{HC_3N}(T)}{k_{HCN}(T)}]  
\label{eq:2}
.\end{eqnarray} 

We assumed the Boltzmann value for the ortho-to-para ratio of H$_2$.
Finally, we note that the disadvantage of GRAPES, namely its inability to
compute the emission separately for the two sources I16293-A and
I16293-B \citep{ja14}, can be neglected here because previous
interferometric observations by \cite{cha05} and \cite{Jorgensen:2011}
have demonstrated that the cyanopolyyne line emission arises from
source I16293-A, as also found by the analysis by \cite{ca11}.

We ran large grids of models varying the four parameters, $X_{in}$,
$X_{out}$, $\alpha,$ and $T_{jump}$, and found the best fit to the
observed fluxes.  In general, we explored the $X_{in}$--$X_{out}$
parameter space by running 10x10 and 20x20 grids for $\alpha$ equal to
-1, 0, +1, and +2, and $T_{jump}$ from 10 to 200 K in steps of 10
K. We note that we first started with a range of 3 or 4 orders of magnitude
in $X_{in}$ and $X_{out}$ to find a first approximate solution, and
then we fine-tuned the grid around it.

\subsection{Results}\label{sec:results}
\subsubsection{HC$_3$N}\label{result:hc3n}

\begin{figure*}[tb]
  \centering
  \includegraphics[width=10cm,angle=-90]{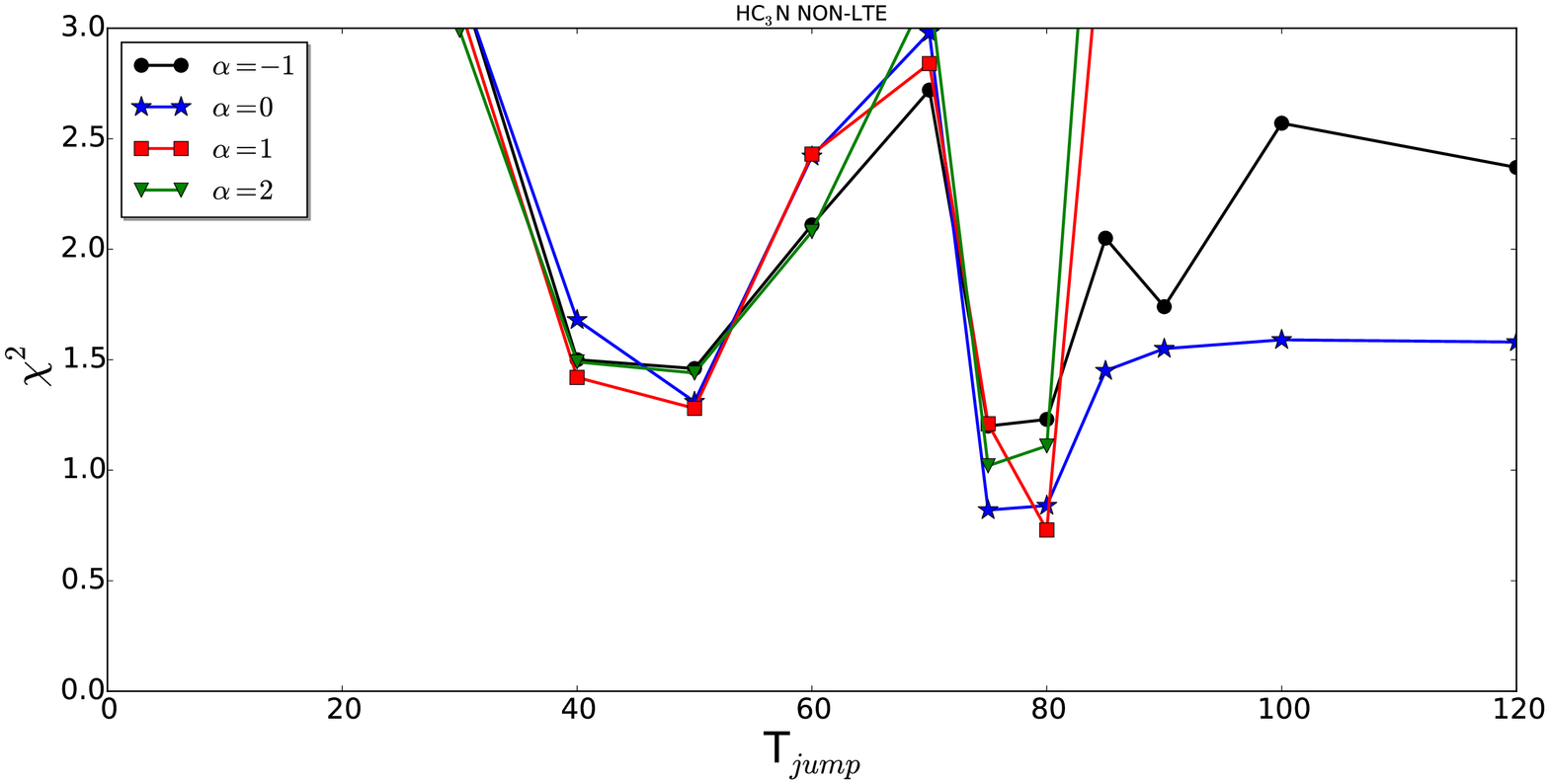}
  \caption{Results of the HC$_3$N modelling. The best reduced $\chi^2$
    optimised with respect to $X_{in}$ and $X_{out}$ as a function of
    $T_{jump}$.}
    \label{fig:hc3n_chi2}
\end{figure*}

The HC$_3$N SLED has been  analysed in two steps, as follows.

\noindent {\it Step 1}: We first ran a grid of models varying
T$_{jump}$, $X_{in}$ , and $X_{out}$ with $\alpha$=0. The best fit for
this first step was obtained with T$_{jump}$=80 K,
$X_{in}=3.6\times10^{-10}$, and $X_{out}=6.0\times10^{-11}$. However,
the fit is not very good (reduced $\chi^2$=1.7). We plotted for each line the
predicted velocity-integrated flux emitted from a shell at a radius
$r$ (namely $dF/dr*r$) as a function of the radius, as shown in
Fig.~\ref{fig:hc3n_contam}. For the three lines with the lowest J (from
J=9 to 11) the predicted shell-flux increases with the radius and
abruptly stops at the maximum radius of the envelope. This means
that these
three lines are very likely contaminated by the molecular cloud, which
could explain the poor $\chi^2$. In the second modelling step, we therefore
excluded these lines and repeated the modelling, as described below.

%
\begin{figure*}
  \centering
  \includegraphics[width=10cm,angle=-90]{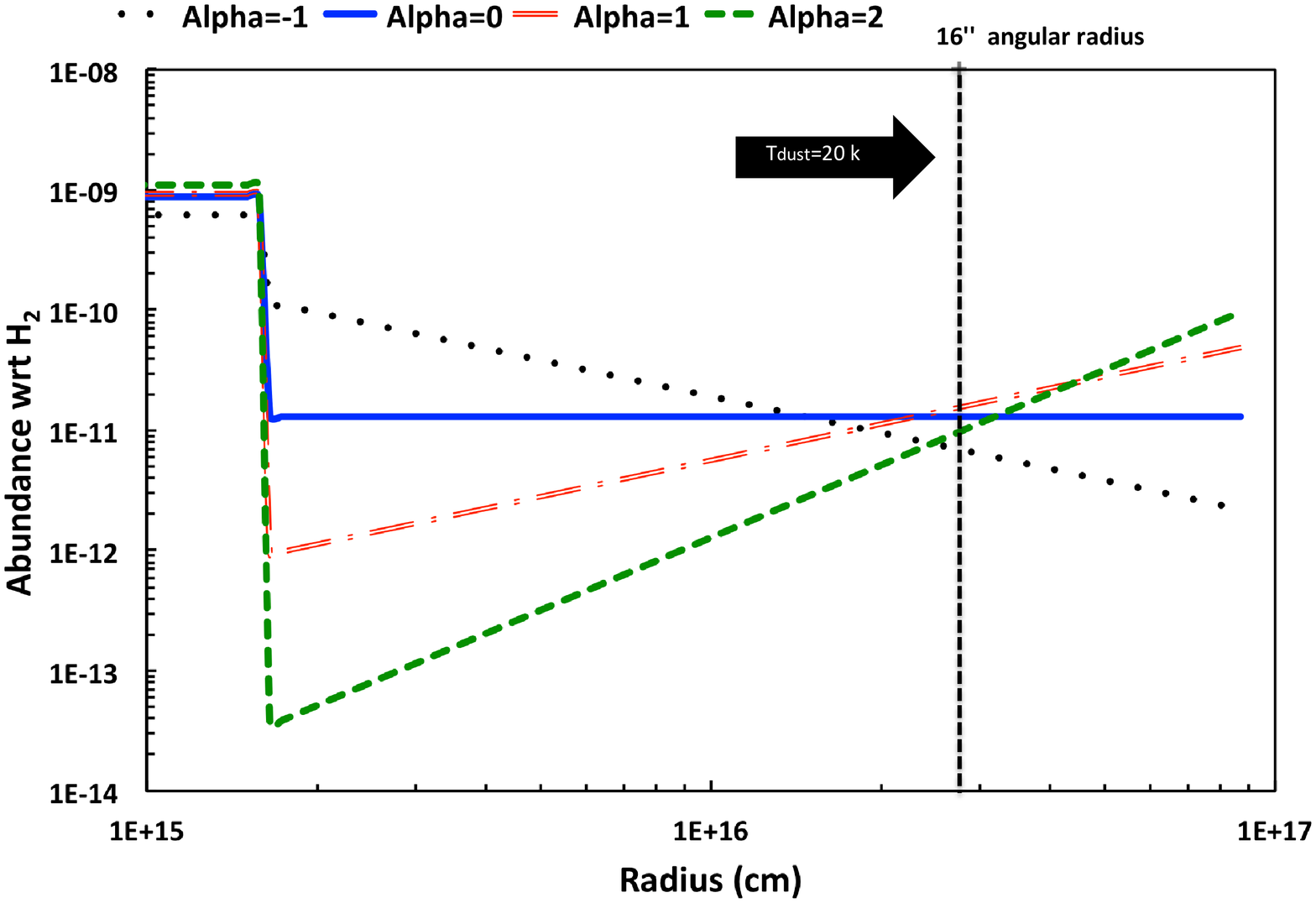}
  \caption{Abundance profiles of the four HC$_3$N best-fit models of
    Table \ref{tab:hc3n_result}. }
    \label{fig:abun}
\end{figure*}

\begin{table*}[tb]
  \centering
  \caption{Results of the HC$_3$N modelling. Values of the best fit
    using four different values of $\alpha$.}
  \begin{tabular}{ccccccccc}
    \hline
Model No.&$\alpha$&T$_{jump}$&X$_{in}$&X$_{out}$&X$_{T20}$ &X$_{in}$/X$_{out}$&X$_{in}$/X$_{T20}$&$\chi^2$\\   
         &      &    [K]   &[$10^{-10}$]& [$10^{-10}$] & [$10^{-10}$]& &  &  \\
          \hline\hline
Model 1& -1 & 80$\pm5$ &6$\pm1$   &0.020$^{+0.009} _{-0.001}$ &0.07$^{+0.03}_{-0.01}$ &300$^{+70}_{-130}$ &90$^{+30}_{-40}$  &1.2\\
Model 2& 0  & 80$\pm5$ &9$\pm1$   &0.13$^{+0.03}_{-0.04}$    &0.13$^{+0.03}_{-0.04}$  &90$^{+20}_{-40}$   &90$^{+20}_{-40}$  &0.8\\
Model 3& 1  & 80$\pm5$ &9$\pm1$   &0.5$^{+0.1} _{-0.2}$      &0.2$\pm0.1$            &18$^{+15}_{-5}$    &45$^{+55}_{-20}$  &0.7\\
Model 4& 2  & 80$\pm5$ &11$\pm1$  &1.0$^{+0.1} _{-0.4}$        &0.10$^{+0.01}_{-0.04}$  &11$^{+9}_{-2}$     &110$^{+90}_{-20}$ &1.1 \\   
    \hline 
  \end{tabular}
  \label{tab:hc3n_result}
\end{table*}

\noindent {\it Step 2}: We ran a grid of models as in step 1 and found a better $\chi^2$ (0.8). We therefore extended the
analysis by varying the $\alpha$ parameter between -1 and +2, together
with the other three parameters T$_{jump}$, $X_{in}$ , and
$X_{out}$. The results are reported in Fig.~\ref{fig:hc3n_chi2} and
Table \ref{tab:hc3n_result}. We note that in Table
\ref{tab:hc3n_result}, we also report the values of the abundance
of the envelope, $X_{T20}$, at a radius equal to the angular size of
the beam with the lowest frequencies, namely $29".5$ in diameter. At
this radius, which is 2.75$\times10^{16}$ cm (around 1800 AU), the envelope dust
temperature is 20 K.

Figure~\ref{fig:hc3n_chi2} shows the $\chi^2$ obtained for each value of
$\alpha$, minimised with respect to $X_{in}$ and X$_{out}$, as a
function of T$_{jump}$.  A $\chi^2$ lower than unity is obtained with
T$_{jump}$ equal to 80 K and $\alpha$ equal to 0 and 1 (Table
\ref{tab:hc3n_result}). Figure~\ref{fig:abun} shows the abundance
profiles of HC$_3$N as predicted by the best-fit models. We note that the
four models have the same T$_{jump}$ (80 K), and the same $X_{in}$
($9\times10^{-10}$) and $X_{T20}$ at 20 K ($\sim1\times10^{-11}$),
within a factor 2. Therefore, the determination of T$_{jump}$, $X_{in}$ , and
$X_{out}$ at 20 K is very robust and depends very little on the
assumption of the HC$_3$N abundance distribution.

Given the similarity of our results regardless of $\alpha$, in the
following we only consider the case of $\alpha$=0.
Figure~\ref{fig:chi2} shows the $\chi^2$ contour plots as a function of
the inner $X_{in}$ and outer $X_{out}$ with $T_{jump}$=80 K.
Figure~\ref{fig:ratio} reports the ratio of the observed over predicted
line fluxes as a function of the upper level energy of the transition,
to show the goodness of the best-fit model.  Finally,
Fig.~\ref{fig:flux} shows the predicted shell-flux as a function of
the radius for a sample of lines.  As a final remark, we note that we
also ran LTE models and obtained approximately the same
results, within 20\% for $X_{in}$ and $X_{out}$ and the same
$T_{jump}$ and $\alpha$.

\subsubsection{HC$_5$N}

Because we now had fewer lines, we decided to assume that HC$_5$N follows the
spatial distribution of HC$_3$N and ran a grid of models with
$\alpha$=0, and T=80 K and $X_{in}$ and $X_{out}$ as free
parameters. The $X_{in}$--$X_{out}$ $\chi^2$ surface is shown in
Fig. \ref{fig:chi2} and the ratio between the observed and best-fit
predicted intensities is shown in Fig.~\ref{fig:ratio}.  Table
\ref{tab:results} summarises the best-fit values.  We only obtained an
upper limit to the $X_{in}$ abundance, $\leq8\times10^{-11}$, while
the $X_{out}$ abundance is $\sim 1\times10^{-11}$. When compared to
HC$_3$N, the HC$_5$N abundance is therefore more than ten times
lower in the hot corino region, while it is the same in the outer
cold envelope.

\subsubsection{DC$_3$N}
Following the discussion on the HC$_3$N line analysis, we did not
consider the three DC$_3$N lines with the lowest upper level energy,
as they are likely contaminated by the molecular cloud.

For the remaining three lines, we adopted the same strategy as for
HC$_5$N for the SLED analysis, namely we adopted $\alpha$=0, and T=80
K and varied $X_{in}$ and $X_{out}$.  The results are shown in
Figs. \ref{fig:chi2} and \ref{fig:ratio}, and summarised in Table
\ref{tab:results}.  We obtained an upper limit of $X_{in}
\leq 4\times10^{-11}$, while $X_{out}$ is $\leq 5\times10^{-12}$.
This implies a deuteration ratio of $\leq 5$\% and $50$\% in the hot
corino and cold envelope, respectively.

\subsubsection{Undetected species and final remarks on the observations}
As mentioned earlier, we did not detect larger cyanopolyynes or
$^{13}$C isotopologues. For the undetected species we derived the
upper limits to the abundance by assuming that the non-detected
emission arises in the hot corino (with a temperature of 80 K,
diameter = $2\arcsec$, N(H$_2$) = 1.5$\times10^{23}$ cm$^{-2}$, and line FWHM =
6 km/s), and cold envelope (with a temperature of 20 K, a diameter =
$30"$, N(H$_2$) = 3.5$\times10^{22}$ cm$^{-2}$, and line FWHM = 3 km/s),
respectively.  The results are listed in Table \ref{tab:results}.

\begin{table*}[!htbp]
\begin{threeparttable}
\centering
\caption{Results of the analysis.}
\begin{tabular}{llccccc}
\hline
Species & Formula  & X$_{in}$ & X$_{T20}$ &  \multicolumn{2}{c} {X/HC$_3$N}&$\chi^2$ \\
 &                  &[$10^{-10}$] & [$10^{-10}$] & [In]  & [T20]& \\
\hline
\multicolumn{7}{c}{Detected cyanopolyynes}\\ \hline

Cyanoacetylene            &HC$_3$N   & 9$\pm1$   & 0.13$^{+0.03}_{-0.04}$   &1          & 1                    & 0.8 \\
Cyanodiacetylene          &HC$_5$N   & $\la $0.8 & 0.110$\pm{0.005}$       &$\la $0.1  & 0.8$^{+0.2}_{-0.4}$ & 1.1 \\
Deuterated cyanoacetylene &DC$_3$N   & $\la $0.4 & 0.049$^{+0.002}_{-0.003}$&$\la $0.04 & 0.4$^{+0.1}_{-0.2}$  & 1.9  \\
\hline

\multicolumn{7}{c}{Undetected cyanopolyynes}\\ \hline
Ethynylisocyanide        & HCCNC                   & $\la 7.5$  & $\la 0.10$ \\
3-Imino-1,2-propa-dienylidene& HNCCC               & $\la 1.2$  & $\la 0.01$ \\
Cyanoacetylene, $^{13}$C   &HCCC-13-N     &$\la 4.5$   &$\la 0.10$  \\
Cyanoacetylene, $^{13}$C     &HCC-13-CN     &$\la 4.5$   &$\la 0.09$  \\
Cyanoacetylene, $^{13}$C     &HC-13-CCN     &$\la 7.5$   &$\la 0.15$   \\
Cyanoacetylene, $^{15}$N    & HCCCN-15               &$\la 1.5$   &$\la 0.15$   \\
                          & DNCCC                  &$\la 1.5$   &$\la 0.04$  \\
Cyanodiacetylene, $^{13}$C & HCCCCC-13-N          &$\la 6.0$   &$\la 0.60$   \\
Cyanodiacetylene, $^{13}$C & HCCCC-13-CN          &$\la 7.5$   &$\la 0.60$   \\
Cyanodiacetylene, $^{13}$C & HCCC-13-CCN          &$\la 4.5$   &$\la 0.45$   \\
Cyanodiacetylene, $^{13}$C & HCC-13-CCCN          &$\la 7.5$   &$\la 0.75$   \\
Cyanodiacetylene, $^{13}$C & HC-13-CCCCN          &$\la 6.0$   &$\la 0.60$   \\
Cyanodiacetylene, $^{15}$N & HCCCCCN-15           &$\la 7.5$   &$\la 0.75$   \\
Cyanodiacetylene, D & DCCCCCN        &$\la 7.5$   &$\la 0.75$   \\
Cyanohexatriyne    & HC$_7$N                       &$\la 15$  &$\la 15$    \\
\hline \hline
\end{tabular}
\begin{tablenotes} 
\item Note: The first two columns report the species name and
  formula. Third and fourth columns report the values of the inner
  abundance $X_{in}$ and the abundances at T$_{dust}$=20 K
  $X_{T20}$. Columns 5 and 6 report the values of the abundance ratios
  DC$_3$N/HC$_3$N and HC$_5$N/HC$_3$N in the inner region and where
  T$_{dust}$=20 K, respectively. The last column reports the $\chi^2$ for
  the best fit of HC$_5$N and DC$_3$N (note that in this last case the
  $\chi^2$ is not reduced).  The top half table lists the detected
  species, the bottom half table the upper limits to the abundance of
  undetected cyanopolyynes (see text, Sect.~ \ref{sec:modeling}).
  \label{tab:results}
  \end{tablenotes}
  \end{threeparttable}
\end{table*}

From the HC$_5$N and DC$_3$N modelling, we conclude that the detected
lines of both species are dominated by the emission from the cold
outer envelope, as we only obtain upper limits to their abundance in
the warm inner region. This conclusion is coherent with the line
widths reported in Table \ref{tab:all}. The HC$_3$N lines (see
also Fig. \ref{fig:hc3n_compatible}) show a clear separation between
the low-energy transitions, for which the FWHM is around 3 km/s, and
the high-energy transitions, for which the FWHM is around 6 km/s or
higher. This difference strongly suggests that the two sets
of lines probe different regions. As the low-energy lines correspond
to E$_{up}$ values lower than 80 K, a very natural explanation is that
they are dominated by emission from the cold envelope, whereas the
high-energy lines (E$_{up} \geq$ 110 K) are dominated by emission from
the inner warm region. As the HC$_5$N and DC$_3$N lines have
  a FWHM of about 3 km/s, it is not surprising that our modelling
concludes that they are dominated by emission from the cold envelope.

\section{Chemical origin of HC$_3$N}\label{sec:chemical-modeling}

As discussed in the Introduction, HC$_3$N is an ubiquitous molecule in
the ISM. There are several routes by which a high HC$_3$N abundance
may arise. Briefly, HC$_3$N can form through the following neutral-neutral
reactions \citep[e.g.][]{Wakelam:2015}:\\
\ce{C2H2 + CN -> HC3N + H }\\
\ce{C4H + N -> HC3N + C}\\
\ce{C3H2 + N -> HC3N + H}\\
\ce{C2H + HNC -> HC3N + H}\\
under typical dense-cloud conditions. Roughly, the first
  reaction contributes to about 80--90\% of the formation of HC$_3$N,
  with the other three swapping in importance depending on the
  initial conditions, but together contributing the remaining 20\%.
Its main destruction channels are either through reactions with He$^+$ or
reactions with atomic carbon, both contributing 30-40\% to the
  destruction, depending on the initial conditions assumed. In dark
clouds (n$_H$ $\sim$ 10$^4$--10$^5$ cm$^{-3}$), it is well known that
HC$_3$N is abundant, together with other carbon chain molecules
\citep[e.g.][]{Suzuki:1992,Caselli:1998}.  Thus, based on its
routes of formation and destruction, we would expect the HC$_3$N
abundance to increase as a function of gas density, at least before
freeze-out onto the dust grains takes over.  Interestingly, however,
the HC$_3$N abundance with respect to H$_2$ estimated in both the
envelope and in the hot corino of IRAS16293 is rather low ($\sim$
10$^{-11}$ and $\sim$ 10$^{-9}$ , respectively). In this section we
therefore discuss the possible physical conditions that may lead to
such low abundances of HC$_3$N.

\subsection{Cold envelope}
First of all, we investigated qualitatively which conditions
in the cold envelope may lead to a fractional abundance of HC$_3$N as
low as 10$^{-11}$. We would like to emphasise that this is difficult
because very small changes in the main atomic elements (O, C, N) or
molecular species (CO) can lead to very large changes in the
abundances of tracers species. Hence, it is likely that small
uncertainties in the initial elemental abundances of the cloud as well
as its depletion history (which is dependent on the time the gas
remains at a particular gas density) can easily lead to differences in
the abundances of species such as HC$_3$N of orders of magnitude.

For our analysis, we made use of the time-dependent chemical model
UCL\_CHEM~\citep{Viti:2004}. The code was used in its simplest form,
namely we only included gas-phase chemistry with no dynamics (the gas
was always kept at a constant density and temperature).  The gas-phase
chemical network is based on the UMIST database~\citep{McElroy:2013}
augmented with updates from the KIDA database as well as new rate
coefficients for the HC$_3$N and HC$_5$N network as estimated
by~\cite{Loison:2014}.  The chemical evolution was followed until
chemical equilibrium was reached.  To understand the chemical
origin of the HC$_3$N abundance that we measured in the cold envelope
of IRAS16293 ($\sim 10^{-11}$), we modelled a gas density of
$2\times10^6$ cm$^{-3}$ and a gas temperature of 20 K, which are the
values inferred by Crimier et al. (2010) at a radius equivalent to the
observations telescope beam (see Sect. \ref{result:hc3n}).

We ran a small grid of models where we varied
\begin{itemize}
\item the carbon and oxygen elemental abundances in the range of
  0.3--1 and 1--3 $\times$10$^{-4}$, respectively;
\item the nitrogen elemental abundance: 6 (solar) and 2 $\times$10$^{-5}$;
\item the cosmic-ray ionisation rate: 5$\times$10$^{-17}$ s$^{-1}$
(assumed as our standard value) and 10 times higher.
\end{itemize}
We note that varying the C, O, and N elemental abundances can be
considered approximately as mimicking qualitatively the degree of
depletion of these elements onto the grain mantles. Furthermore, we
varied the cosmic-ray ionisation rate because it is a rather
uncertain parameter that could be enhanced with respect to the
standard value because of the presence of X-rays or an energetic
$\geq$MeV particles source embedded inside the envelope
\citep[e.g.][]{Doty:2004,Ceccarelli:2014}.

\begin{figure}[tb]
  \centering
   \includegraphics[width=6cm,angle=-90]{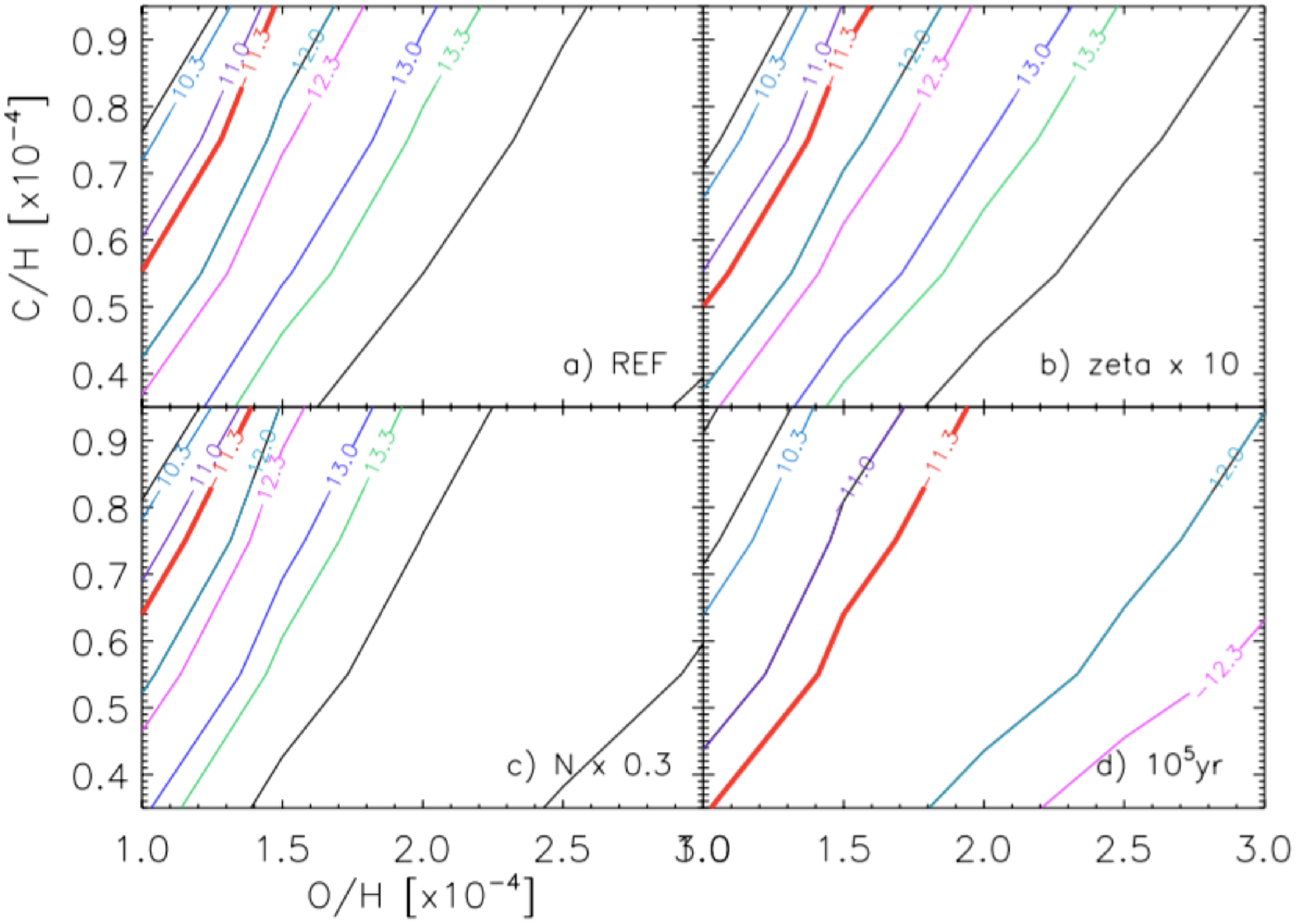}

  \caption{Predicted HC$_3$N abundance (in log) as a function of the
    O/H (x-axis) and C/H (y-axis) for four cases: the reference model,
    described in the text (upper left panel), and then the same, but
    with a cosmic-ray ionisation rate increased by a factor ten (upper
    right panel), a nitrogen elemental abundance decreased by a factor
    three (lower left panel) and at a time of 10$^5$yr (lower right
    panel). The thick red lines mark the HC$_3$N abundance measured
    in the cold envelope of IRAS16293.}
    \label{fig:hc3n_chemi}
\end{figure}
The results of the modelling are plotted in
Fig.~\ref{fig:hc3n_chemi}. The top left panel is our reference
model, with a standard cosmic-ray ionisation rate (5$\times$10$^{-17}$
s$^{-1}$), at chemical equilibrium ($\geq$ 10$^6$ yr), and with a
solar abundance of nitrogen (6.2$\times$10$^{-5}$). The
values of the HC$_3$N abundances are log of the fractional abundance
with respect to the total number of hydrogen nuclei, hence the best
match with the observations is for a value of -11.3.
We find that this value is reached within our reference model for low
values of O/H ($\leq 1.5\times 10^{-4}$) and for a O/C ratio between
1.5 and 2, with the lowest O/C ratio needed with the highest O/H
abundance. This implies that both oxygen and carbon are mostly frozen
onto the grain icy mantles by about the same factor, as the O/C ratio
is similar to the one in the Sun \citep[1.8; see e.g. ][]{Asplund:2009}
and in the HII regions \citep[1.4; see e.g.][]{Garcia-Rojas:2007}.
Increasing the cosmic-ray ionisation rate by a factor of ten
changes the above conclusions little, having as an effect a
slightly larger parameter space in O/H and O/C to reproduce the
observed HC$_3$N abundance. In contrast, decreasing N/H by a
factor of three slightly diminishes the O/H and O/C parameter space.
Finally, the largest O/H and O/C parameter space that reproduces the
observed HC$_3$N abundance in the cold envelope of IRAS16293 is
obtained by considering earlier times, 10$^5$ years (bottom right of
Fig.~\ref{fig:hc3n_chemi}). This possibly implies that the gas in the envelope is at a similar
age to that of the protostar. 
 
\subsection{Hot corino}
As we move towards the centre of the protostar(s), a jump in the
HC$_3$N abundance by roughly two orders of magnitude is observed at
around 80 K. This means that the first question to answer is what this 80 K
represents. According to the temperature programmed desorption
(TPD)
experiments by~\cite{Collings:2004} and successive similar works,
in general, the ice sublimation is a complex process and does not
occur at one single temperature, but in several steps. In the
particular case of iced HC$_3$N, it is expected to have two
sublimation peaks, at a dust temperature of about 80 and 100 K,
respectively \citep[for details, see][]{Viti:2004}. The first
corresponds to the so-called {\it \textup{volcano}} ice desorption, and the
second corresponds to the ice co-desorption, namely the whole ice
sublimation. Therefore, our {\it \textup{measured}} jump temperature of 80 K
indicates that the most important sublimation is due to the volcano
desorption, while the co-desorption injects back only a minor fraction of
the frozen HC$_3$N. In addition, the 80 K jump also tells us that any other processes contributing to the formation of HC$_3$N
at lower ($\leq80$ K) temperatures have to be negligible. In other words,
considering the reactions in the synthesis of HC$_3$N reported at the
beginning of Sect. \ref{sec:chemical-modeling}, the desorption of
C-bearing iced species, for example CH$_4$, has to provide a
negligible contribution to the HC$_3$N abundance. 

To understand what all this implies, we ran UCL\_CHEM again,
but this time we included freeze-out of the gas species in the cold
phase to simulate the cold and dense prestellar phase.  We then
allowed thermal evaporation of the icy mantles as a function of the
species and temperature of the dust following the recipe
of~\cite{Collings:2004}, in which the ice sublimation occurs in
several steps \citep[for details, see][]{Viti:2004}.
We find that the jump in the HC$_3$N abundance from 10$^{-11}$ to
10$^{-9}$ occurs only if thermal evaporation due to the volcano peak
occurs quickly, on a timescale of $\leq 10^{3-4}$yr, implying that the
increase in the dust temperature to $\sim$ 80 K must also occur
quickly, on a similar timescale.
In addition, if frozen species containing C, such as CO and CH$_4$,
sublimate at earlier times, namely at colder temperatures, and remain
in the gas for too long before the volcano explosion, then the
HC$_3$N abundance reaches much too high abundances too quickly. In
other words, the sublimation of these species must have occurred not
much before the volcano sublimation, again on timescales of $\leq10^3$
yr.

\begin{figure}[bt]
  \centering
  \includegraphics[width=7cm,angle=-90]{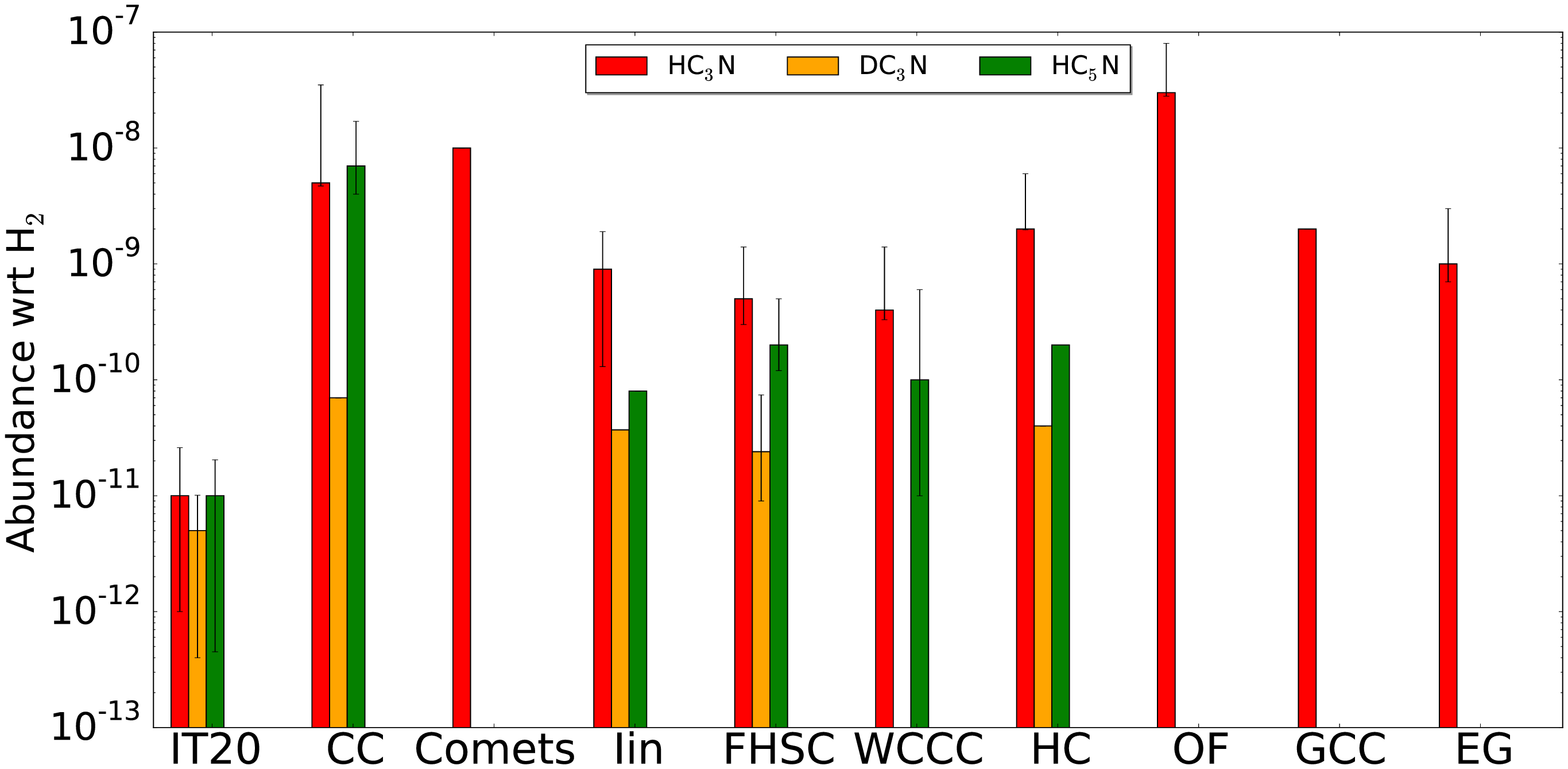}

  \caption{Abundances of cyanopolynes in different sources: IRAS16293
    outer envelope (IT20) and inner region (Iin) (this work), cold
    clouds (CC)~\citep{wi96,mi14}, comet Hale-Bopp at 1 AU assuming
    H$_2$O/H$_2$=5$\times10^{-5}$ (Comets)~\citep{bo20}, first
    hydrostatic core sources (FHSC)~\citep{co12}, warm carbon-chain
    chemistry sources (WCCC)~\citep{sa08,jo04}, massive hot
    cores (HC)~\citep{sc02,es13}, outflow sources (OF)~\citep{ba97,sc02},
    Galactic Centre clouds (GCC)~\citep{ma93,al11}, and external
    galaxies (EG) ~\citep{al11}. }
  \label{fig:comp}
\end{figure}

\subsection{HC$_5$N}
We finally note that none of our models can reproduce an HC$_5$N
abundance comparable to the one of HC$_3$N, as observed in the cold
envelope of IRAS16293. All models predict a much lower HC$_5$N
abundance, by more than a factor ten, than the HC$_3$N
abundance. Since in the warm (T$\geq$80 K) region we only derive
an upper limit for the HC$_5$N abundance (which is at least ten times
lower than that of HC$_3$N), the models cannot be constrained. 
The failure of our models to reproduce HC$_5$N in the cold envelope is
most likely due to a lack of a comprehensive network for the formation and
destruction of the species HC$_5$N and therefore calls for a
revision of its gaseous chemistry.
Here we briefly summarize the main routes of formation and destruction for this species  in our models: \\
\ce{C2H2 + C3N -> HC5N + H }\\
\ce{C4H + HNC -> HC5N + H}\\
contributing to its formation by $\sim$ 70\% and 30\%, respectively, and\\
\ce{H3+ + HC5N -> HC5NH+ + H2}\\
\ce{HCO+ + HC5N -> HC5NH+ + CO}\\
contributing to its destruction by $\sim$ 50\% and 40\%, respectively.

It is clear that since both HC$_3$N and HC$_5$N are tracer species (with abundances $\leq 10^{-9}$),
at least in this particular source, a
more complete network is required, possibly including cyanopolynes with a higher
number of carbon than included here, together with a
proper treatment of the gas-grain chemistry that is needed to
properly model the cold envelope. This will be the scope of future
work.

\section{Discussion}\label{sec:duscussion}

\subsection{General remarks on cyanopolyynes in different environments}

In the Introduction, we mentioned that cyanopolyynes are almost
ubiquitous in the ISM. Here we compare the cyanopolyyne abundance
derived in this work with that found in various Galactic and
extragalactic environments that possess different conditions
(temperature, density, and history).  Figure~\ref{fig:comp} graphically
shows this comparison. From this figure, we note the following: 
(i) HC$_3$N is present everywhere in the ISM with
relatively high abundances; (ii) the abundance in the cold envelope of
IRAS16293 is the lowest in the plot, implying a high degree of
freezing of oxygen and carbon, as indeed suggested by the chemical
model analysis; (iii) HC$_3$N and HC$_5$N have relatively similar
abundances in the IRAS16293 outer envelope, cold cloud, first
hydrostatic core (FHSC), and WCCC
sources, namely in the cold objects of the figure, implying that the
derived ratio HC$_5$N/HC$_3$N$\sim1$ in the IRAS16293 cold envelope is
not, after all, a peculiarity, but probably due to the cold temperature;
and finally, (iv) DC$_3$N is only detected in cold objects
except in the warm envelope of IRAS16293 and the hot cores, suggesting
that in these sources, it is linked to the mantle sublimation, in one way or another, as again found by our chemical modelling.

\begin{figure}[tb]
  \centering
  \includegraphics[width=6cm,angle=-90]{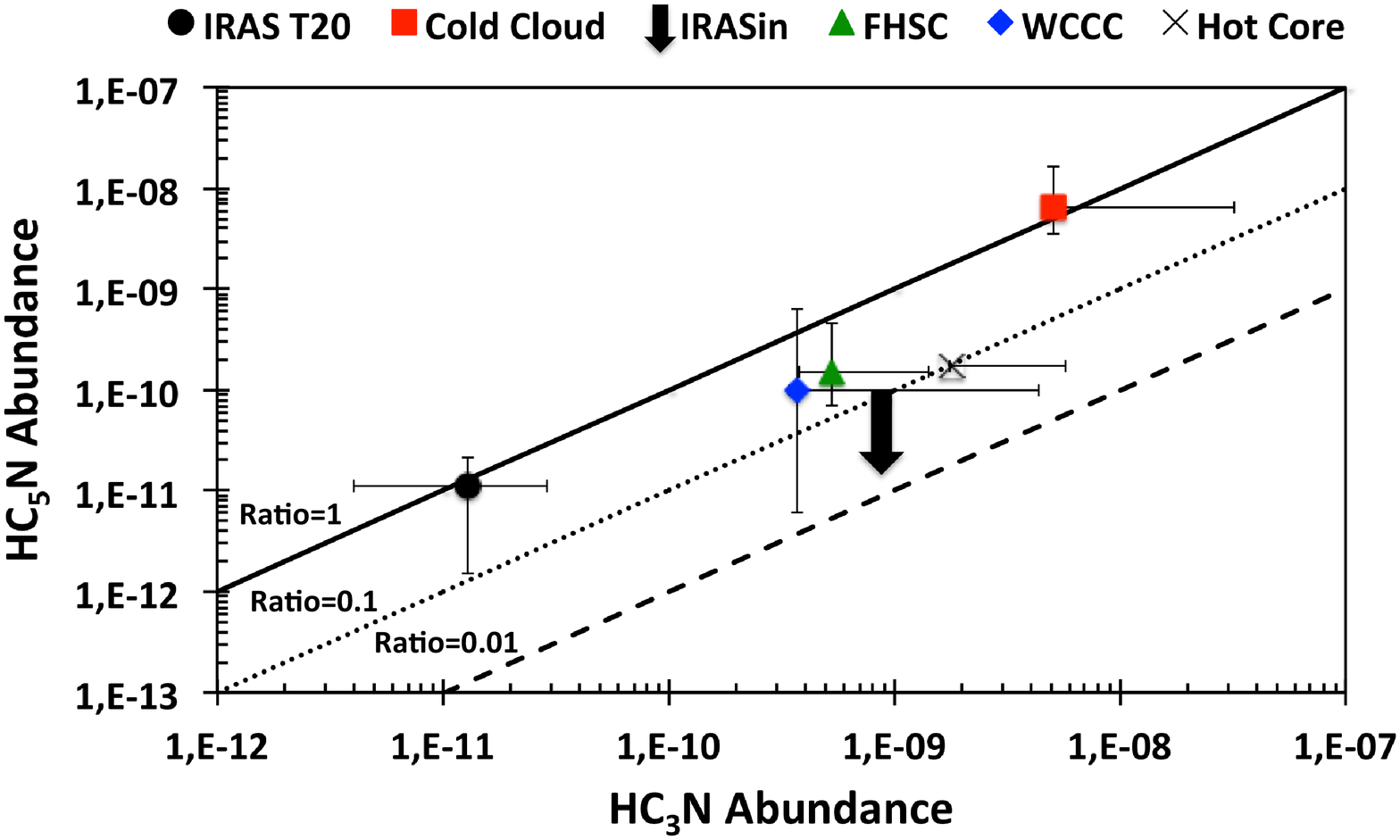}

  \caption{Abundance of HC$_5$N as a function of the abundance of
    HC$_3$N in different protostellar and cold sources: inner (black
    arrow) and outer envelope (T$_{20}$) of IRAS16293 (black filled
    circle) presented in this work, warm carbon-chain chemistry (WCCC)
    sources (blue diamond) \citep{sa08,jo04}, first hydrostatic core
    (FHSC) source (green triangle) \citep{co12}, hot cores sources
    (cross) \citep{sc02,es13}, and Galactic Centre clouds (red square)
    \citep{ma93,al11}.}
  \label{fig:hc5n}
\end{figure}

\subsection{Present and past history of IRAS16293}
Our analysis of the TIMASSS spectral survey reveals the presence of
HC$_3$N, HC$_5$N, and DC$_3$N in IRAS16293. 
HC$_3$N was previously detected by~\cite{van-Dishoeck:1993}, and its
abundance has been analysed by~\cite{sc02} with a two-step
abundance model, where the jump was assumed to occur at 90 K. ~\cite{sc02}
 found HC$_3$N/H$_2 \sim 10^{-9}$ in the inner warm part
and $\leq \sim 10^{-10}$ in the outer envelope. We note that they were unable
to derive a value for the outer envelope or estimate where the jump
occurs because they only detected three lines. Nonetheless, their
estimates are in excellent agreement with our new estimates (Table
\ref{tab:results}). 

A very important point is that we were not only able to estimate the cold
envelope abundance, but also to determine where the jump occurs,
an essential information for the determination of the origin of HC$_3$N in
IRAS16293. Our chemical modeling (Sect. ~\ref{sec:chemical-modeling})
showed that in the cold envelope, the HC$_3$N abundance is reproduced
for low values of oxygen and carbon in the gas phase, implying that
these species are heavily frozen onto the grain mantles. The O/C ratio
is between 1.5 and 2, similar to solar, if the chemistry of
the envelope is a fossil, namely built up during the $\sim10^{7}$ yr
life of the parental molecular cloud. If, in contrast, the
chemistry was reset (for any reason) and evolved in a shorter
time, for instance in $10^5$ yr, the O/C ratio that reproduces the
observations could be as high as 3, but the oxygen and carbon have
anyway remained mostly frozen onto the grain mantles. Probably the
most important point to remark is that the HC$_3$N abundance is very
low, a very tiny fraction of the CO abundance, the main reservoir of
the carbon, so that little variation in the CO abundance results in
large variation in the HC$_3$N abundance.

A second extremely interesting point is that the HC$_3$N abundance
undergoes a jump of about one hundred when the dust temperature
reaches 80 K. These two values provide us with very strong constraints
on how the collapse of IRAS16293 occurred: it must have occurred so
fast that the sublimation of C-bearing ices, such as CO and methane,
has not yet produced HC$_3$N in enough large abundances to mask
the jump that is due to the volcano sublimation of the ices at 80
K~\citep{Collings:2004,Viti:2004}.  An approximate value of this time
is $\sim10^3$ yr, based on our modelling. In other words, the HC$_3$N
measured abundance jump temperature and abundance in the warm envelope
both suggest that IRAS16293 is a very young object, as expected,
and that the envelope dust heating took no more than $\sim10^3$ yr to
occur.

A third result from the present study is the large measured abundance
of HC$_5$N, for the first time detected in a solar-type protostar. It
is indeed only ten times lower than HC$_3$N in the inner warm region
and similar to the HC$_3$N abundance in the outer cold envelope. 
When compared to other protostellar sources where HC$_5$N has been
detected, this ratio is not so anomalous. Figure
\ref{fig:hc5n} shows the two abundances in several protostellar
sources and cold clouds. The two coldest sources have both an
HC$_5$N/HC$_3$N$\sim 1$, while in the other sources this ratio is
$\sim 0.1$. 
However, our chemical model fails to reproduce such a high abundance
of HC$_5$N, even though we included part of the most updated chemical network,
recently revised by~\cite{Loison:2014}. Evidently, we are missing
key reactions that form this species.
\begin{figure}[tb]
  \centering
  \includegraphics[width=7cm,angle=-90]{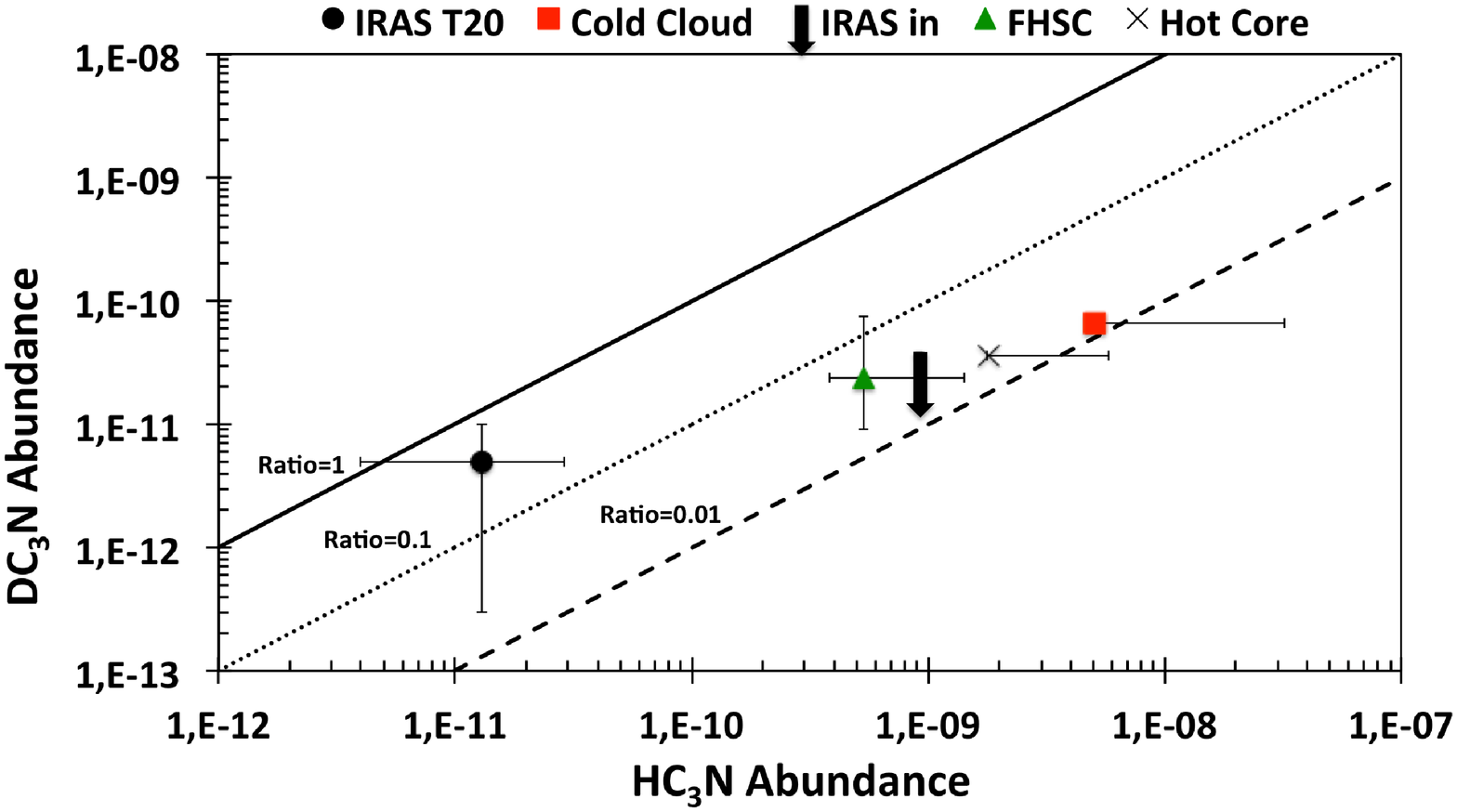}

  \caption{Abundance of DC$_3$N as a function of the abundance of
    HC$_3$N in different protostellar and cold sources. The symbols
    are the same as those in Fig. \ref{fig:hc5n}.}
    \label{fig:dc3n}
\end{figure}

\subsection{HC$_3$N deuteration}
Finally, we detected for the first time the DC$_3$N in a solar-type
protostar.  The deuteration of HC$_3$N is about 50\% in the outer cold
envelope and lower than 5\% in the warm part. This also provides us
with important clues on the present and past history of IRAS16293. 
First, the high deuteration in the cold envelope tells us that this
is a present-day product, namely it is caused by a cold and
CO-depleted gas, in agreement with previous observations and
theoretical predictions \citep[e.g.][]{Ce14}.

The very low deuteration in the warm part is, in contrast, more
intriguing. It cannot be the result of an initial high
deuteration that has been diminished by gas-phase reactions, because,
as we have argued above, HC$_3$N is the result of the volcano
sublimation, which occurred because of a quick, $\leq10^3$yr, heating
of the dust. This means that it must be a fossil from the dust before the warming
phase.  This is probably one of the few very clear cases of fossil
deuteration where there are no doubts that it is pristine (another
one, to our knowledge, is the one of HDCO in the protostellar
molecular shock L1157-B1; \cite{Fontani:2014}). The deuteration in
IRAS16293 is very high in general, with a doubly
deuterated ratio of formaldehyde of $\sim30$\% , for example
\citep{Ceccarelli:1998},
and a triply deuterated ratio of methanol of a few
percent~\citep{Parise:2004}.  While at present we still do not have a
clear measure of the formaldehyde deuteration in the warmer and colder
envelope separately \cite[e.g.][]{Ceccarelli:2001}, methanol should be
entirely concentrated in the warm region. However, recently,
deuterated formamide (NH$_2$CDO and NHDCHO) has been detected
by ALMA in IRAS16293, and the observations clearly show that formamide
line emission is associated with the warm hot
corino~\citep{Coutens:2016}. The measured deuteration ratio is a few
percent, similar to that found in HC$_3$N in the present
work. Therefore, the molecular deuteration is a complex phenomenon
even within the same source.

If our reasoning above is correct, namely the present-day gaseous
HC$_3$N is the sublimation product of previously frozen HC$_3$N, then
its relatively low deuteration tells us that this species is an
early chemical product, namely it was formed in a time when the
gas temperature was not too low and, mostly importantly, the CO was
not depleted yet \citep[see e.g.][]{Ce14}. This is in perfect
agreement with the models of HC$_3$N formation, which show
that HC$_3$N is {\it \textup{abundantly}} formed {\it \textup{before}} the full trapping
of carbon into CO \citep[see e.g.][and Sect. 5]{Loison:2014}. Therefore,
the low DC$_3$N/HC$_3$N in the warm part is well explained by the
picture that abundant HC$_3$N has been formed during the tenuous
molecular cloud phase and then frozen onto the ice mantles when the
condensation that gave birth to IRAS16293 increased in density while
it decreased in temperature. Then, we predict that the deuteration of
HC$_3$N in the hot corino regions could be a good probe of the
timescale of the collapse as well. Of course, detailed modelling will be
necessary to quantify this prediction.

Meanwhile, the collection of previous measures of DC$_3$N/HC$_3$N,
shown in Figure \ref{fig:dc3n}, provides support to our thesis.
With the exception of the cold envelope of IRAS16293, all other
sources where DC$_3$N has been observed, possess a HC$_3$N deuteration
of a few percent, in agreement with the hypothesis that HC$_3$N is an
early chemical product.

\section{Conclusions}\label{sec:conclusions}

We detected several lines from cyanoacetylene (HC$_3$N) and
cyanodiacetylene (HC$_5$N), and provided an upper limit to the
abundance of cyanotriacetylene (HC$_7$N) and other undetected
cyanopolyynes.  We also reported the first detection of deuterated
cyanoacetylene, DC$_3$N, in a solar-type protostar. In contrast,
we did not detect any $^{13}$C cyanopolyyne isotopologue. We found
that the HC$_3$N abundance is roughly
constant($\sim 1.3\times10^{-11}$) in the outer cold envelope of
IRAS16293-2422 and it increases, as a step-function, by about a factor
100 in the inner region where the dust temperature exceeds 80 K. The
HC$_5$N has an abundance similar to HC$_3$N in the outer envelope and
about a factor of ten lower in the inner region.

A comparison with a  chemical model provided constraints on the oxygen
and  carbon gaseous abundance in the outer envelope and, most
importantly, on the age of the source. The HC$_3$N abundance derived
in the inner region and where the jump occurs also provided strong
constraints on the time taken for the dust to warm up to 80 K, which
has to be less than $\sim 10^3-10^4$ yr. 

Finally, the cyanoacetylene deuteration is about 50\% in the outer
envelope and $\sim 5$\% in the warm inner region. The relatively low
deuteration in the warm region suggests that we are seeing an almost
pristine fossil of the HC$_3$N, abundantly formed in the tenuous phase
of the pre-collapse and then frozen into the grain mantles at a later
phase.

\begin{acknowledgements} 
  This research was supported in part by the National Science
  Foundation under Grant No. NSF PHY11-25915. We acknowledge the
  financial support from the university of Al-Muthana and ministry of
  higher education and scientific research in Iraq. EM acknowledges
  support from the Brazilian agency FAPESP under grants
  2014/22095-6 and 2015/22254-0.
\end{acknowledgements}


\begin{thebibliography}{59}
\expandafter\ifx\csname natexlab\endcsname\relax\def\natexlab#1{#1}\fi

\bibitem[{{Aladro} {et~al.}(2011){Aladro}, {Mart{\'{\i}}n-Pintado},
  {Mart{\'{\i}}n}, {Mauersberger}, \& {Bayet}}]{al11}
{Aladro}, R., {Mart{\'{\i}}n-Pintado}, J., {Mart{\'{\i}}n}, S., {Mauersberger},
  R., \& {Bayet}, E. 2011, \aap, 525, A89

\bibitem[{{Asplund} {et~al.}(2009){Asplund}, {Grevesse}, {Sauval}, \&
  {Scott}}]{Asplund:2009}
{Asplund}, M., {Grevesse}, N., {Sauval}, A.~J., \& {Scott}, P. 2009, \araa, 47,
  481

\bibitem[{{Bachiller} \& {P{\'e}rez Guti{\'e}rrez}(1997)}]{ba97}
{Bachiller}, R. \& {P{\'e}rez Guti{\'e}rrez}, M. 1997, \apjl, 487, L93

\bibitem[{{Bell} {et~al.}(1997){Bell}, {Feldman}, {Travers}, {McCarthy},
  {Gottlieb}, \& {Thaddeus}}]{be97}
{Bell}, M.~B., {Feldman}, P.~A., {Travers}, M.~J., {et~al.} 1997, \apjl, 483,
  L61

\bibitem[{{Ben Abdallah} {et~al.}(2012){Ben Abdallah}, {Najar}, {Jaidane},
  {Dumouchel}, \& {Lique}}]{Ben-Abdallah:2012}
{Ben Abdallah}, D., {Najar}, F., {Jaidane}, N., {Dumouchel}, F., \& {Lique}, F.
  2012, \mnras, 419, 2441

\bibitem[{{Bockel{\'e}e-Morvan} {et~al.}(2000){Bockel{\'e}e-Morvan}, {Lis},
  {Wink}, {Despois}, {Crovisier}, {Bachiller}, {Benford}, {Biver}, {Colom},
  {Davies}, {G{\'e}rard}, {Germain}, {Houde}, {Mehringer}, {Moreno}, {Paubert},
  {Phillips}, \& {Rauer}}]{bo20}
{Bockel{\'e}e-Morvan}, D., {Lis}, D.~C., {Wink}, J.~E., {et~al.} 2000, \aap,
  353, 1101

\bibitem[{Brack(1998)}]{bra98}
Brack, A. 1998, The Molecular Origins of Life: Assembling Pieces of the Puzzle
  (Cambridge University Press)

\bibitem[{{Caselli} \& {Ceccarelli}(2012)}]{Caselli:2012}
{Caselli}, P. \& {Ceccarelli}, C. 2012, \aapr, 20, 56

\bibitem[{{Caselli} {et~al.}(1998){Caselli}, {Walmsley}, {Terzieva}, \&
  {Herbst}}]{Caselli:1998}
{Caselli}, P., {Walmsley}, C.~M., {Terzieva}, R., \& {Herbst}, E. 1998, \apj,
  499, 234

\bibitem[{{Caux} {et~al.}(2011){Caux}, {Kahane}, {Castets}, {Coutens},
  {Ceccarelli}, {Bacmann}, {Bisschop}, {Bottinelli}, {Comito}, {Helmich},
  {Lefloch}, {Parise}, {Schilke}, {Tielens}, {van Dishoeck}, {Vastel},
  {Wakelam}, \& {Walters}}]{ca11}
{Caux}, E., {Kahane}, C., {Castets}, A., {et~al.} 2011, \aap, 532, A23

\bibitem[{{Cazaux} {et~al.}(2003){Cazaux}, {Tielens}, {Ceccarelli}, {Castets},
  {Wakelam}, {Caux}, {Parise}, \& {Teyssier}}]{caz03}
{Cazaux}, S., {Tielens}, A.~G.~G.~M., {Ceccarelli}, C., {et~al.} 2003, \apjl,
  593, L51

\bibitem[{{Ceccarelli} {et~al.}(2014{\natexlab{a}}){Ceccarelli}, {Caselli},
  {Bockel{\'e}e-Morvan}, {Mousis}, {Pizzarello}, {Robert}, \& {Semenov}}]{Ce14}
{Ceccarelli}, C., {Caselli}, P., {Bockel{\'e}e-Morvan}, D., {et~al.}
  2014{\natexlab{a}}, Protostars and Planets VI, 859

\bibitem[{{Ceccarelli} {et~al.}(2000){Ceccarelli}, {Castets}, {Caux},
  {Hollenbach}, {Loinard}, {Molinari}, \& {Tielens}}]{cec00}
{Ceccarelli}, C., {Castets}, A., {Caux}, E., {et~al.} 2000, \aap, 355, 1129

\bibitem[{{Ceccarelli} {et~al.}(1998){Ceccarelli}, {Castets}, {Loinard},
  {Caux}, \& {Tielens}}]{Ceccarelli:1998}
{Ceccarelli}, C., {Castets}, A., {Loinard}, L., {Caux}, E., \& {Tielens},
  A.~G.~G.~M. 1998, \aap, 338, L43

\bibitem[{{Ceccarelli} {et~al.}(2014{\natexlab{b}}){Ceccarelli}, {Dominik},
  {L{\'o}pez-Sepulcre}, {Kama}, {Padovani}, {Caux}, \&
  {Caselli}}]{Ceccarelli:2014}
{Ceccarelli}, C., {Dominik}, C., {L{\'o}pez-Sepulcre}, A., {et~al.}
  2014{\natexlab{b}}, \apjl, 790, L1

\bibitem[{{Ceccarelli} {et~al.}(1996){Ceccarelli}, {Hollenbach}, \&
  {Tielens}}]{ce96}
{Ceccarelli}, C., {Hollenbach}, D.~J., \& {Tielens}, A.~G.~G.~M. 1996, \apj,
  471, 400

\bibitem[{{Ceccarelli} {et~al.}(2001){Ceccarelli}, {Loinard}, {Castets},
  {Tielens}, {Caux}, {Lefloch}, \& {Vastel}}]{Ceccarelli:2001}
{Ceccarelli}, C., {Loinard}, L., {Castets}, A., {et~al.} 2001, \aap, 372, 998

\bibitem[{{Ceccarelli} {et~al.}(2003){Ceccarelli}, {Maret}, {Tielens},
  {Castets}, \& {Caux}}]{ce03}
{Ceccarelli}, C., {Maret}, S., {Tielens}, A.~G.~G.~M., {Castets}, A., \&
  {Caux}, E. 2003, \aap, 410, 587

\bibitem[{{Cernicharo} \& {Guelin}(1996)}]{cer96}
{Cernicharo}, J. \& {Guelin}, M. 1996, \aap, 309, L27

\bibitem[{{Chandler} {et~al.}(2005){Chandler}, {Brogan}, {Shirley}, \&
  {Loinard}}]{cha05}
{Chandler}, C.~J., {Brogan}, C.~L., {Shirley}, Y.~L., \& {Loinard}, L. 2005,
  \apj, 632, 371

\bibitem[{{Clarke} \& {Ferris}(1995)}]{Clarke:1995}
{Clarke}, D.~W. \& {Ferris}, J.~P. 1995, \icarus, 115, 119

\bibitem[{{Collings} {et~al.}(2004){Collings}, {Anderson}, {Chen}, {Dever},
  {Viti}, {Williams}, \& {McCoustra}}]{Collings:2004}
{Collings}, M.~P., {Anderson}, M.~A., {Chen}, R., {et~al.} 2004, \mnras, 354,
  1133

\bibitem[{{Cordiner} {et~al.}(2012){Cordiner}, {Charnley}, {Wirstr{\"o}m}, \&
  {Smith}}]{co12}
{Cordiner}, M.~A., {Charnley}, S.~B., {Wirstr{\"o}m}, E.~S., \& {Smith}, R.~G.
  2012, \apj, 744, 131

\bibitem[{{Coutens} {et~al.}(2016){Coutens}, {J{\o}rgensen}, {van der Wiel},
  {M{\"u}ller}, {Lykke}, {Bjerkeli}, {Bourke}, {Calcutt}, {Drozdovskaya},
  {Favre}, {Fayolle}, {Garrod}, {Jacobsen}, {Ligterink}, {{\"O}berg},
  {Persson}, {van Dishoeck}, \& {Wampfler}}]{Coutens:2016}
{Coutens}, A., {J{\o}rgensen}, J.~K., {van der Wiel}, M.~H.~D., {et~al.} 2016,
  \aap, 590, L6

\bibitem[{{Crimier} {et~al.}(2010){Crimier}, {Ceccarelli}, {Maret},
  {Bottinelli}, {Caux}, {Kahane}, {Lis}, \& {Olofsson}}]{cr10}
{Crimier}, N., {Ceccarelli}, C., {Maret}, S., {et~al.} 2010, \aap, 519, A65

\bibitem[{{Doty} {et~al.}(2004){Doty}, {Sch{\"o}ier}, \& {van
  Dishoeck}}]{Doty:2004}
{Doty}, S.~D., {Sch{\"o}ier}, F.~L., \& {van Dishoeck}, E.~F. 2004, \aap, 418,
  1021

\bibitem[{{Esplugues} {et~al.}(2013){Esplugues}, {Cernicharo}, {Viti},
  {Goicoechea}, {Tercero}, {Marcelino}, {Palau}, {Bell}, {Bergin}, {Crockett},
  \& {Wang}}]{es13}
{Esplugues}, G.~B., {Cernicharo}, J., {Viti}, S., {et~al.} 2013, \aap, 559, A51

\bibitem[{{Faure} {et~al.}(2016){Faure}, {Lique}, \& {Wiesenfeld}}]{Faure:2016}
{Faure}, A., {Lique}, F., \& {Wiesenfeld}, L. 2016, \mnras, 460, 2103

\bibitem[{{Fontani} {et~al.}(2014){Fontani}, {Codella}, {Ceccarelli},
  {Lefloch}, {Viti}, \& {Benedettini}}]{Fontani:2014}
{Fontani}, F., {Codella}, C., {Ceccarelli}, C., {et~al.} 2014, \apjl, 788, L43

\bibitem[{{Friesen} {et~al.}(2013){Friesen}, {Medeiros}, {Schnee}, {Bourke},
  {Francesco}, {Gutermuth}, \& {Myers}}]{fr13}
{Friesen}, R.~K., {Medeiros}, L., {Schnee}, S., {et~al.} 2013, \mnras, 436,
  1513

\bibitem[{{Garc{\'{\i}}a-Rojas} \& {Esteban}(2007)}]{Garcia-Rojas:2007}
{Garc{\'{\i}}a-Rojas}, J. \& {Esteban}, C. 2007, \apj, 670, 457

\bibitem[{Goesmann {et~al.}(2015)Goesmann, Rosenbauer, Bredeh{\"o}ft, Cabane,
  Ehrenfreund, Gautier, Giri, Kr{\"u}ger, Le~Roy, MacDermott, McKenna-Lawlor,
  Meierhenrich, Caro, Raulin, Roll, Steele, Steininger, Sternberg, Szopa,
  Thiemann, \& Ulamec}]{goe15}
Goesmann, F., Rosenbauer, H., Bredeh{\"o}ft, J.~H., {et~al.} 2015, Science, 349

\bibitem[{{Jaber} {et~al.}(2014){Jaber}, {Ceccarelli}, {Kahane}, \&
  {Caux}}]{ja14}
{Jaber}, A.~A., {Ceccarelli}, C., {Kahane}, C., \& {Caux}, E. 2014, \apj, 791,
  29

\bibitem[{{J{\o}rgensen} {et~al.}(2011){J{\o}rgensen}, {Bourke}, {Nguyen
  Luong}, \& {Takakuwa}}]{Jorgensen:2011}
{J{\o}rgensen}, J.~K., {Bourke}, T.~L., {Nguyen Luong}, Q., \& {Takakuwa}, S.
  2011, \aap, 534, A100

\bibitem[{{J{\o}rgensen} {et~al.}(2004){J{\o}rgensen}, {Sch{\"o}ier}, \& {van
  Dishoeck}}]{jo04}
{J{\o}rgensen}, J.~K., {Sch{\"o}ier}, F.~L., \& {van Dishoeck}, E.~F. 2004,
  \aap, 416, 603

\bibitem[{{Loinard} {et~al.}(2008){Loinard}, {Torres}, {Mioduszewski}, \&
  {Rodr{\'{\i}}guez}}]{lo08}
{Loinard}, L., {Torres}, R.~M., {Mioduszewski}, A.~J., \& {Rodr{\'{\i}}guez},
  L.~F. 2008, \apjl, 675, L29

\bibitem[{{Loinard} {et~al.}(2013){Loinard}, {Zapata}, {Rodr{\'{\i}}guez},
  {Pech}, {Chandler}, {Brogan}, {Wilner}, {Ho}, {Parise}, {Hartmann}, {Zhu},
  {Takahashi}, \& {Trejo}}]{lo13}
{Loinard}, L., {Zapata}, L.~A., {Rodr{\'{\i}}guez}, L.~F., {et~al.} 2013,
  \mnras, 430, L10

\bibitem[{{Loison} {et~al.}(2014){Loison}, {Wakelam}, {Hickson}, {Bergeat}, \&
  {Mereau}}]{Loison:2014}
{Loison}, J.-C., {Wakelam}, V., {Hickson}, K.~M., {Bergeat}, A., \& {Mereau},
  R. 2014, \mnras, 437, 930

\bibitem[{{Lunine}(2009)}]{lun09}
{Lunine}. 2009, EPJ Web of Conferences, 1, 267

\bibitem[{{Marr} {et~al.}(1993){Marr}, {Wright}, \& {Backer}}]{ma93}
{Marr}, J.~M., {Wright}, M.~C.~H., \& {Backer}, D.~C. 1993, \apj, 411, 667

\bibitem[{{McElroy} {et~al.}(2013){McElroy}, {Walsh}, {Markwick}, {Cordiner},
  {Smith}, \& {Millar}}]{McElroy:2013}
{McElroy}, D., {Walsh}, C., {Markwick}, A.~J., {et~al.} 2013, \aap, 550, A36

\bibitem[{{Miettinen}(2014)}]{mi14}
{Miettinen}, O. 2014, \aap, 562, A3

\bibitem[{{Mizuno} {et~al.}(1990){Mizuno}, {Fukui}, {Iwata}, {Nozawa}, \&
  {Takano}}]{mi90}
{Mizuno}, A., {Fukui}, Y., {Iwata}, T., {Nozawa}, S., \& {Takano}, T. 1990,
  \apj, 356, 184

\bibitem[{{M{\"u}ller} {et~al.}(2005){M{\"u}ller}, {Schl{\"o}der}, {Stutzki},
  \& {Winnewisser}}]{mu05}
{M{\"u}ller}, H.~S.~P., {Schl{\"o}der}, F., {Stutzki}, J., \& {Winnewisser}, G.
  2005, Journal of Molecular Structure, 742, 215

\bibitem[{{Mundy} {et~al.}(1992){Mundy}, {Wootten}, {Wilking}, {Blake}, \&
  {Sargent}}]{mu92}
{Mundy}, L.~G., {Wootten}, A., {Wilking}, B.~A., {Blake}, G.~A., \& {Sargent},
  A.~I. 1992, \apj, 385, 306

\bibitem[{{Parise} {et~al.}(2004){Parise}, {Castets}, {Herbst}, {Caux},
  {Ceccarelli}, {Mukhopadhyay}, \& {Tielens}}]{Parise:2004}
{Parise}, B., {Castets}, A., {Herbst}, E., {et~al.} 2004, \aap, 416, 159

\bibitem[{{Pickett} {et~al.}(1998){Pickett}, {Poynter}, {Cohen}, {Delitsky},
  {Pearson}, \& {M{\"u}ller}}]{pi98}
{Pickett}, H.~M., {Poynter}, R.~L., {Cohen}, E.~A., {et~al.} 1998, \jqsrt, 60,
  883

\bibitem[{{Pineda} {et~al.}(2012){Pineda}, {Maury}, {Fuller}, {Testi},
  {Garc{\'{\i}}a-Appadoo}, {Peck}, {Villard}, {Corder}, {van Kempen}, {Turner},
  {Tachihara}, \& {Dent}}]{pi12}
{Pineda}, J.~E., {Maury}, A.~J., {Fuller}, G.~A., {et~al.} 2012, \aap, 544, L7

\bibitem[{{Sakai} {et~al.}(2008){Sakai}, {Sakai}, {Hirota}, \&
  {Yamamoto}}]{sa08}
{Sakai}, N., {Sakai}, T., {Hirota}, T., \& {Yamamoto}, S. 2008, \apj, 672, 371

\bibitem[{{Sch{\"o}ier} {et~al.}(2002){Sch{\"o}ier}, {J{\o}rgensen}, {van
  Dishoeck}, \& {Blake}}]{sc02}
{Sch{\"o}ier}, F.~L., {J{\o}rgensen}, J.~K., {van Dishoeck}, E.~F., \& {Blake},
  G.~A. 2002, \aap, 390, 1001

\bibitem[{{Snell} {et~al.}(1981){Snell}, {Schloerb}, {Young}, {Hjalmarson}, \&
  {Friberg}}]{Snell:1981}
{Snell}, R.~L., {Schloerb}, F.~P., {Young}, J.~S., {Hjalmarson}, A., \&
  {Friberg}, P. 1981, \apj, 244, 45

\bibitem[{{Suzuki} {et~al.}(1992){Suzuki}, {Yamamoto}, {Ohishi}, {Kaifu},
  {Ishikawa}, {Hirahara}, \& {Takano}}]{Suzuki:1992}
{Suzuki}, H., {Yamamoto}, S., {Ohishi}, M., {et~al.} 1992, \apj, 392, 551

\bibitem[{{van Dishoeck} {et~al.}(1993){van Dishoeck}, {Blake}, {Draine}, \&
  {Lunine}}]{van-Dishoeck:1993}
{van Dishoeck}, E.~F., {Blake}, G.~A., {Draine}, B.~T., \& {Lunine}, J.~I.
  1993, in Protostars and Planets III, ed. E.~H. {Levy} \& J.~I. {Lunine},
  163--241

\bibitem[{{Viti} {et~al.}(2004){Viti}, {Collings}, {Dever}, {McCoustra}, \&
  {Williams}}]{Viti:2004}
{Viti}, S., {Collings}, M.~P., {Dever}, J.~W., {McCoustra}, M.~R.~S., \&
  {Williams}, D.~A. 2004, \mnras, 354, 1141

\bibitem[{{Wakelam} {et~al.}(2015){Wakelam}, {Loison}, {Herbst}, {Pavone},
  {Bergeat}, {B{\'e}roff}, {Chabot}, {Faure}, {Galli}, {Geppert}, {Gerlich},
  {Gratier}, {Harada}, {Hickson}, {Honvault}, {Klippenstein}, {Le Picard},
  {Nyman}, {Ruaud}, {Schlemmer}, {Sims}, {Talbi}, {Tennyson}, \&
  {Wester}}]{Wakelam:2015}
{Wakelam}, V., {Loison}, J.-C., {Herbst}, E., {et~al.} 2015, \apjs, 217, 20

\bibitem[{{Wernli} {et~al.}(2007){Wernli}, {Wiesenfeld}, {Faure}, \&
  {Valiron}}]{we07}
{Wernli}, M., {Wiesenfeld}, L., {Faure}, A., \& {Valiron}, P. 2007, \aap, 464,
  1147

\bibitem[{{Winstanley} \& {Nejad}(1996)}]{wi96}
{Winstanley}, N. \& {Nejad}, L.~A.~M. 1996, \apss, 240, 13

\bibitem[{{Wootten}(1989)}]{wo89}
{Wootten}, A. 1989, \apj, 337, 858

\bibitem[{{Zapata} {et~al.}(2013){Zapata}, {Loinard}, {Rodr{\'{\i}}guez},
  {Hern{\'a}ndez-Hern{\'a}ndez}, {Takahashi}, {Trejo}, \& {Parise}}]{za13}
{Zapata}, L.~A., {Loinard}, L., {Rodr{\'{\i}}guez}, L.~F., {et~al.} 2013,
  \apjl, 764, L14

\end{thebibliography}


\appendix
\section{Figures that show the HC$_3$N modelling}

\begin{figure*}[]
  \centering
  \includegraphics[width=10cm]{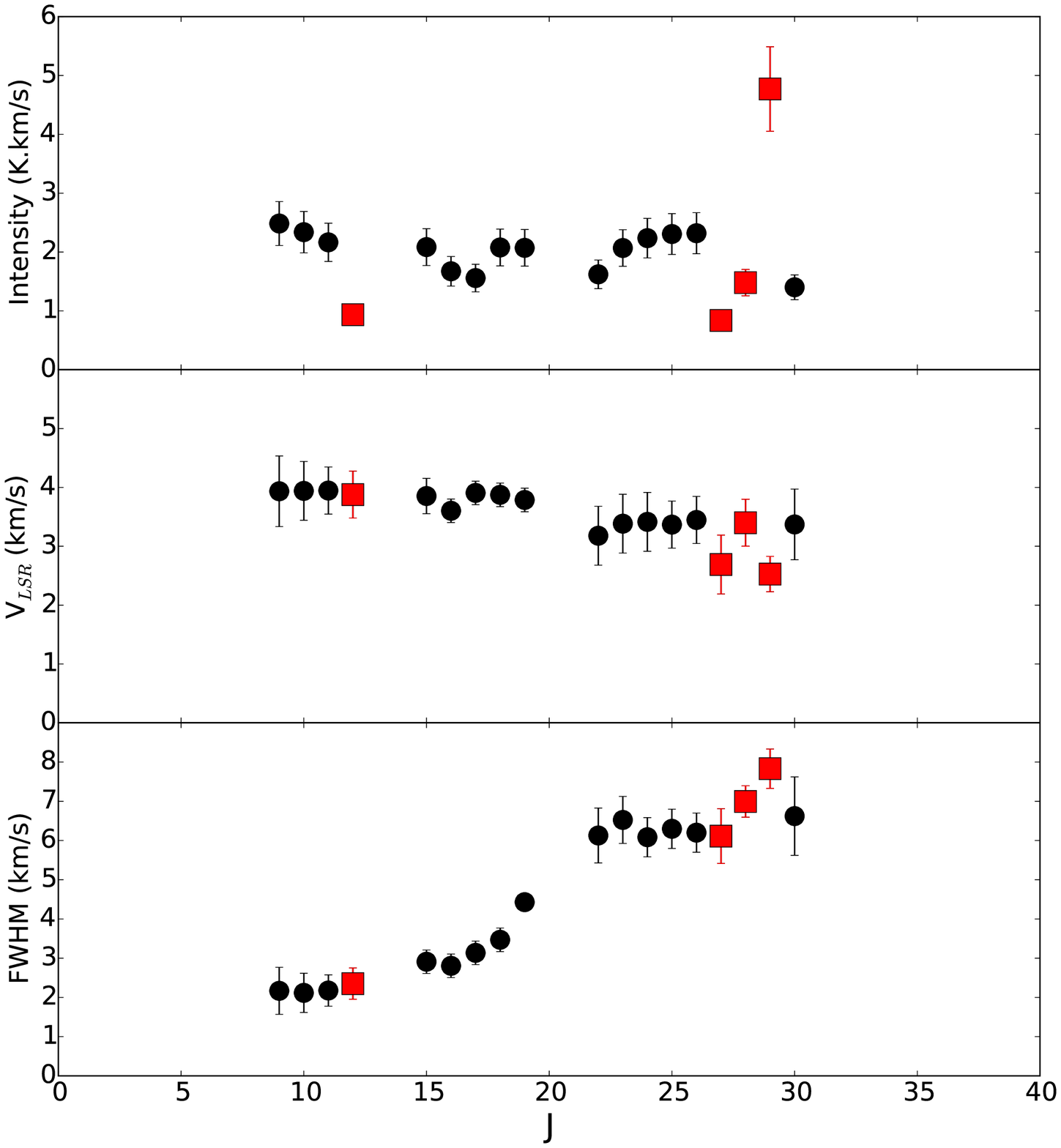}

  \caption{HC$_3$N line intensity (upper panel), rest velocity V$_{LSR}$ (middle panel),
    and FWHM (bottom panel) as a function of the upper J of the
    transition. The red squares show the lines that have been discarded
    because they did not satisfy all criteria 3 to 5 of Sect.
    \ref{subsec:sp_iden} (see text).}
\label{fig:hc3n_compatible} 
\end{figure*}

\begin{figure*}[bt]
  \centering
  \includegraphics[width=12cm,angle=-90]{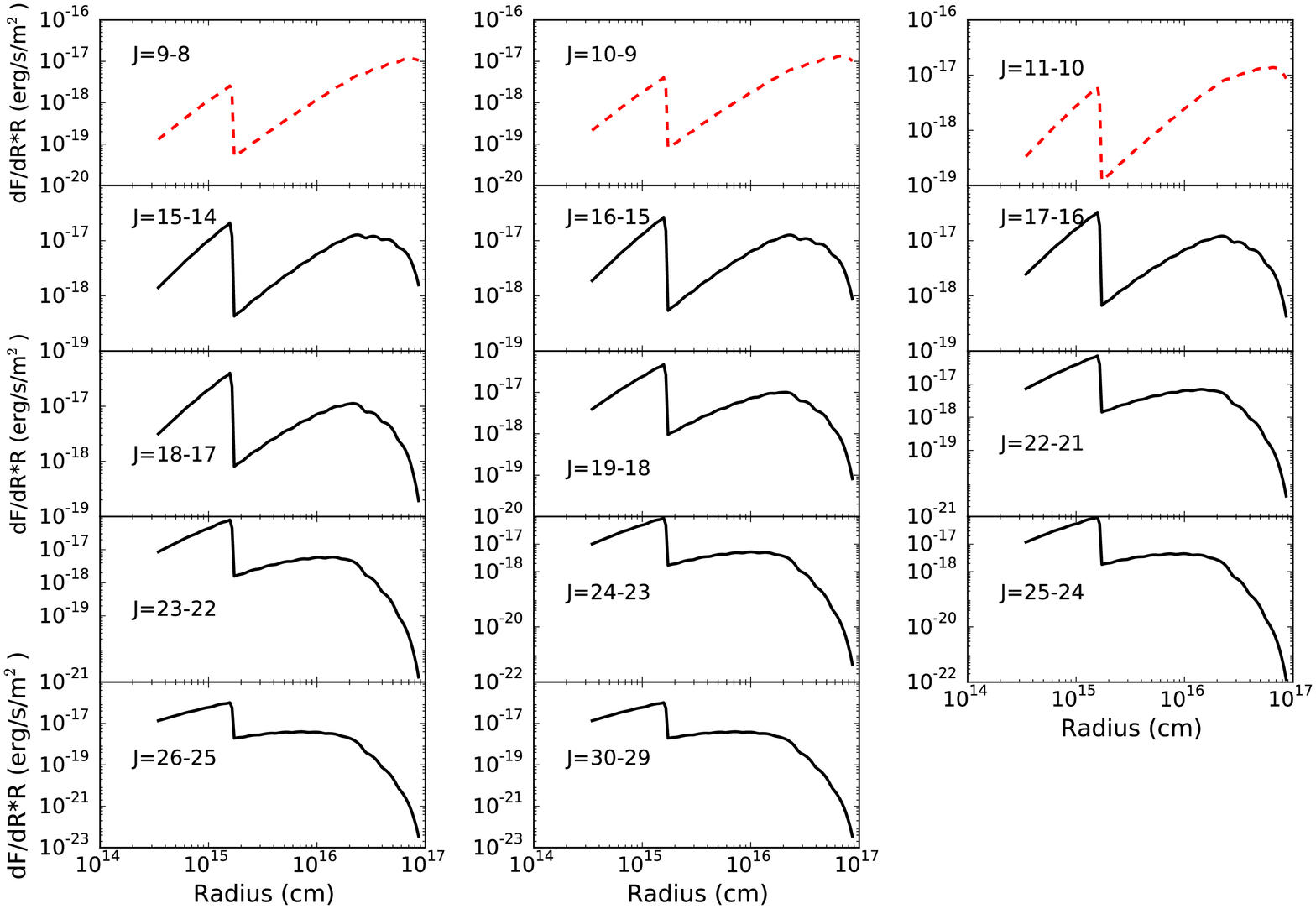}

  \caption{Predicted contribution to the integrated line intensity
    ($dF/dr*r$) of a shell at a radius $r$ for the HC$_3$N lines. This
    model corresponds to $\alpha$=0, T$_{jump}$=80 K,
    $X_{in}=3.6\times10^{-10}$, and $X_{out}=6.0\times10^{-11}$. The three
    upper red dashed curves show an increasing emission towards the
    maximum radius and very likely contaminated by the molecular cloud
    (see Sect. \ref{result:hc3n}). }
  \label{fig:hc3n_contam}
\end{figure*}
%

\begin{figure*}
  \centering 
    \includegraphics[width=12cm,angle=-90]{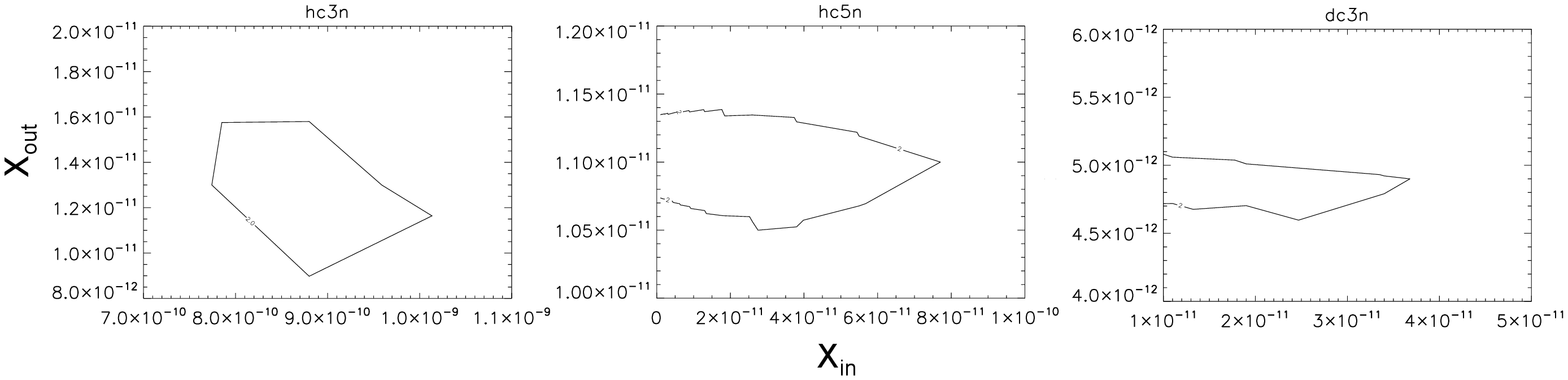}    
  
  \caption{ $\chi^2$ contour plots for HC$_3$N (left),
    HC$_5$N (middle) and DC$_3$N (right) as a function of $X_{in}$ and
    $X_{out}$. The predictions refer to a model with $T_{jump}$=80 K
    and $\alpha$=0.}
    \label{fig:chi2}
\end{figure*}
\begin{figure*}[htp]
  \centering
 \includegraphics[width=10cm,angle=-90]{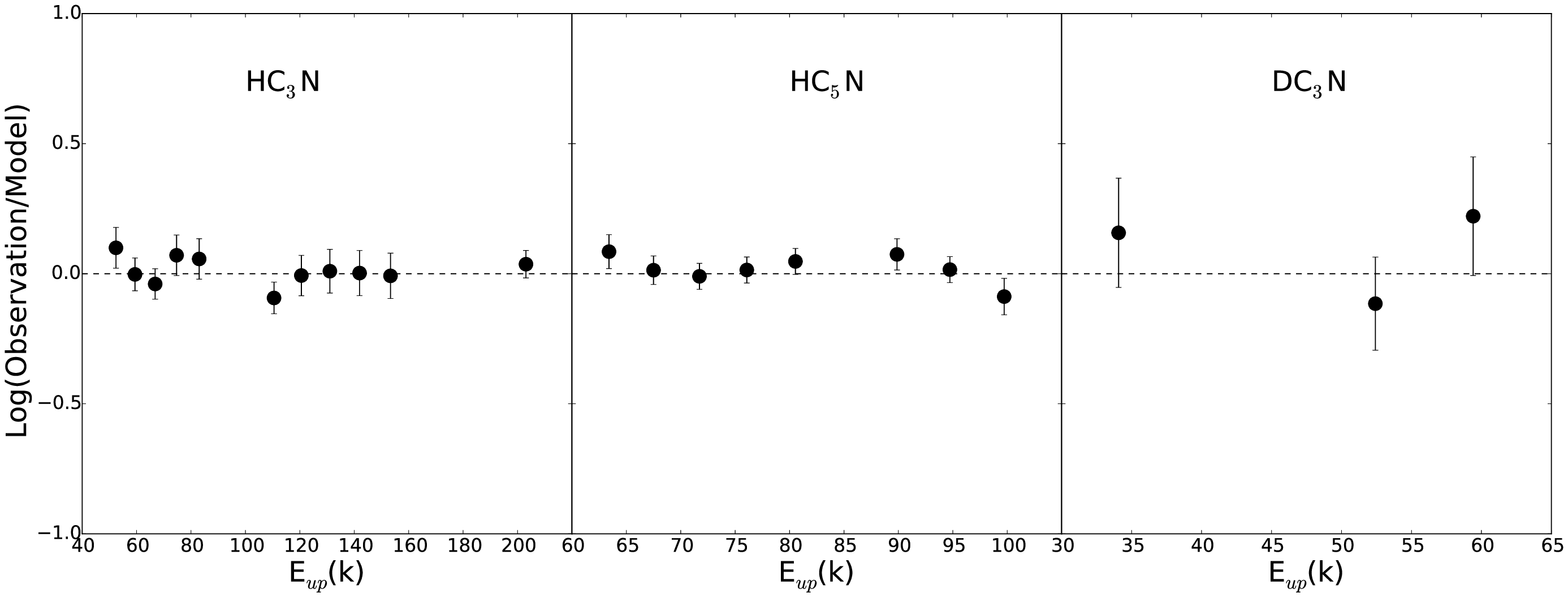}
  \caption{Ratio of the observed over predicted line flux as a
    function of the upper level energy of the transition for the model
    2 ($T_{jump}$=80 K and $\alpha$=0), for HC$_3$N (left),
    HC$_5$N (middle), and DC$_3$N (right), respectively. }
    \label{fig:ratio}
\end{figure*}

\begin{figure*}[htp]
  \centering
 \includegraphics[width=12cm,angle=-90]{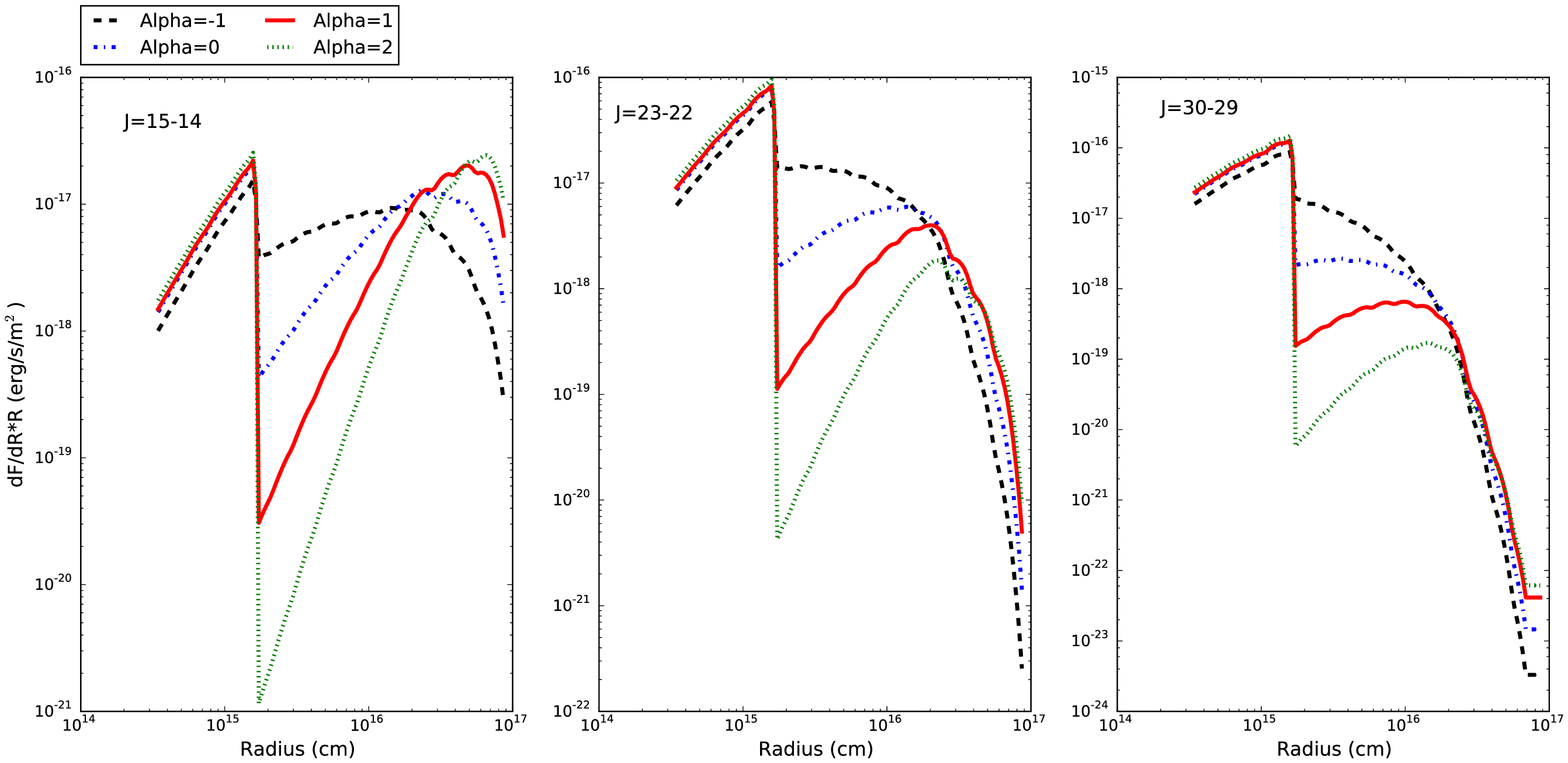}
 
 \caption{Velocity-integrated flux emitted from each shell at a
   radius $r$ ($dF/dr*r$) as a function of the radius for the HC$_3$N
   four best NLTE model, for three low, middle, and high values of
   J. }
    \label{fig:flux}

\end{figure*}

\end{document}